\documentclass{aastex631}
\usepackage{float}
\usepackage{graphicx}
\usepackage{amsmath}
\usepackage{booktabs}
\usepackage{multirow}
\usepackage{textcomp}

\usepackage{placeins}
\usepackage{hyperref}
\usepackage{needspace}

\begin{document}

\title{The Fate of Ice Sheets on Near-Synchronously Rotating Land Planets} 

\author{Sarah Silverman}
\affiliation{\rm Department of Earth and Planetary Sciences, McGill University, Montréal, Canada; sarah.a.silverman@mail.mcgill.ca}
\author{Thomas Navarro}
\affiliation{\rm Space Science Institute 4765 Walnut St, Suite B, Boulder, CO 80301}
\author{Martin Turbet}
\affiliation{\rm Laboratoire de Météorologie Dynamique/IPSL, CNRS, Sorbonne Université, Ecole Normale Supérieure, PSL Research
University, Ecole Polytechnique, 75005 Paris, France}
\affiliation{\rm Laboratoire d’astrophysique de Bordeaux, Univ. Bordeaux, CNRS, B18N, allée Geoffroy Saint-Hilaire, 33615 Pessac,
France}
\author{Nicolas B. Cowan}
\affiliation{\rm Department of Earth and Planetary Sciences, McGill University, Montréal, Canada; sarah.a.silverman@mail.mcgill.ca}
\affiliation{\rm  Department of Physics, McGill University, Montréal, Canada}
\author{Natalya Gomez}
\affiliation{\rm Department of Earth and Planetary Sciences, McGill University, Montréal, Canada; sarah.a.silverman@mail.mcgill.ca}
\author{Siddharth Bhatnagar}
\affiliation{\rm Observatoire astronomique de l’Université de Genève, Versoix, Switzerland}
\affiliation{\rm Centre pour la vie dans l’Univers de l’Université de Genève, Genève, Switzerland}
\affiliation{\rm Group of Applied Physics and Institute for Environmental Sciences, University of Geneva, Geneva, Switzerland}
\author{Emeline Bolmont}
\affiliation{\rm Observatoire astronomique de l’Université de Genève, Versoix, Switzerland}
\affiliation{\rm Centre pour la vie dans l’Univers de l’Université de Genève, Genève, Switzerland}

\begin{abstract}
Rocky exoplanets orbiting M dwarf stars are often assumed to be synchronously rotating, such that water-trapping in ice sheets on the permanent nightside may pose a threat to habitability. However, in compact M dwarf systems, gravitational interactions between planets may cause the substellar point to librate or circulate on millennial timescales and sporadically change direction or speed. Here, we use a 3D global climate model to quantify, for the first time, the water-trapping capabilities of these ``near"\nobreakdash-synchronously rotating exoplanets, with TRAPPIST-1e as our test case. Across a range of synodic day lengths, we find that there is a permanent polar ice reservoir and an increasingly large build-up of lower-latitude transient nightside ice for longer days. A persistent ice-free region develops on the dayside where surface conditions can sustain liquid water. However, this steady-state ice distribution may never be realized since the accumulation timescale may be slower than changes in the spin regime. Additionally, we revise the nightside ice build-up on a synchronously rotating exoplanet using a coupled atmosphere+ice numerical model that accounts for atmospheric lapse rate and isostatic adjustment feedbacks. We find that accounting for this ice-elevation feedback reduces the predicted ice accumulation by a factor of $\sim$3. Taken together, our study suggests that water-trapping in ice sheets is less severe than previously thought for land planets orbiting M dwarf stars.

\end{abstract}


\section{Introduction}

The long-term stability of liquid water is a prerequisite for the emergence and evolution of life as we know it \citep{brack1993liquid, des2008nasa, cockell2016habitability, westall2018importance}. Water is not only required for biological processes, but is implicated in the silicate weathering thermostat that stabilizes climate \citep{walker1981negative, abbot2012indication, alibert2014radius, foley2015role, foley2019habitability}. Hence, surface liquid water is the canonical definition of planetary habitability \citep{abe1993physical, kasting1993habitable}. 

The vast majority of temperate rocky planets orbit in the habitable zone around M dwarf stars, which comprise 75\% of all stars in our Galaxy \citep{henry2006solar, henry2024character}. Many such planets have been discovered orbiting nearby stars: Proxima Centauri b \citep{anglada2016terrestrial}, TRAPPIST-1d, e, f \citep{gillon2016temperate, gillon2017seven}, and LHS 1140b \citep{dittmann2017temperate}. In recent years, the launch of the James Webb Space Telescope has accelerated the characterization of rocky worlds around M dwarfs (see \citealp{kreidberg2025first} for a review). Given their abundance and observability potential, assessing their habitability is central to the search for life beyond the Solar System (see \citealt{shields2016habitability} for a review).

M dwarfs are the lowest mass and least luminous main sequence stars, with a habitable zone much closer-in than for a Sun-like star \citep{tarter2007reappraisal, kopparapu2013habitable, kopparapu2014habitable}. The strong tides raised on planets orbiting so closely should cause them to rotate synchronously, with one hemisphere always facing the star and the other in perpetual darkness \citep{kasting1993habitable, barnes2017tidal, turbet2018modeling}. Consequently, M dwarf exoplanets have been abundantly simulated in this rotational state and have been shown capable of sustaining surface liquid water despite the asymmetric stellar irradiation \citep[e.g.,][]{joshi1997simulations, joshi2003climate, merlis2010atmospheric, wordsworth2011gliese, yang2013stabilizing, hu2014role, turbet2016habitability, bolmont2016habitability, kopparapu2017habitable, haqq2018demarcating, wolf2019simulated, lefevre20213d, sergeev2022bistability, taniguchi2026atmospheric}. 

Synchronously rotating planets nonetheless are at risk of having a substantial portion of their surface water trapped into ice on the nightside ~\citep{heath1999habitability, joshi2003climate, leconte20133d, menou2013water, yang2014water, turbet2016habitability, ojha2022liquid}. The seminal work of ~\cite{menou2013water} showed that hemisphere-sized ice sheets could lock up to the equivalent of 90\% of Earth’s surface water, posing a particular threat to planets with low water inventories relative to Earth: such planets may be left with a desiccated dayside. 

However, this nightside cold-trap scenario depends on synchronous rotation, which is not a guaranteed outcome for exoplanets around M dwarfs. In more eccentric orbits, planets may be trapped into 2:1, 3:2, or higher spin-orbit resonances before tidal synchronization can occur \citep{makarov2012dynamical, rodriguez2012spin, wang2014climate, ribas2016habitability}. These resonant configurations may lead to the formation of thinner ice that is confined to higher latitudes due to more efficient heat redistribution and increased geothermal heat flux from tidal heating \citep{leconte20133d, turbet2016habitability}. 

Of particular relevance to this work is the fact that the majority of M-dwarf stars host compact, multi-planet systems \citep{dressing2013occurrence, dressing2015occurrence, tuomi2019frequency}, where gravitational interactions between neighboring planets may prevent synchronous rotation and spin-orbit resonances altogether \citep{vinson2019chaotic, chen2023sporadic, shakespeare2023day, revol2024drifts}. These interactions perturb the mean motion of the respective planets and cause the substellar point (SSP) to slowly drift along the equator, either librating or fully circulating on decade- up to millennia-long timescales \citep{revol2024drifts}. The time for the SSP to complete a full circulation across the equator defines the length of a synodic day. SSP drift occurs so slowly that these planets are near-synchronously rotating: the planet’s spin rate in an inertial frame, and hence Coriolis forces, large-scale atmospheric circulation, and the day-night temperature contrast, are nearly identical to those on a synchronously-rotating world. In other words, as the sub-stellar point slowly drifts around relative to the geography, the atmosphere relative to the star remains unchanged. \cite{vinson2019chaotic} refer to this SSP drift as ``chaotic'' because planet-planet interactions cause it to reverse direction or change speed every thousand to hundreds of thousands of years. \cite{shakespeare2023day} show that these spin regime changes may occur on even shorter timescales, after just decades or centuries. The drift of starlight over the planetary surface of these near-synchronously rotating exoplanets on long and variable timescales has important implications for where surface ice may grow, recede, and stabilize.

Previous studies of the compact, seven-planet TRAPPIST-1 system have explored the impact of sporadically-changing near-synchronous states on climate, which included examinations of ice formation and the ice-albedo feedback \citep{chen2023sporadic, shakespeare2023day}. \cite{chen2023sporadic} simulated TRAPPIST-1e and f using the 3D global climate model (GCM) ExoPlaSim \citep{paradise2022exoplasim} and found that planet f may be vulnerable to global sea ice coverage, whereas planet~e, receiving higher instellation, can retain a persistent dayside ocean. Additionally, \cite{shakespeare2023day} examined whether increased albedo from dayside ice sheets would make the TRAPPIST-1 planets vulnerable to snowball states by using the 1D energy balance model HEXTOR \citep{haqq2022energy}. In contrast, they found that the weak ice-albedo feedback in the redder TRAPPIST-1 spectrum would prevent such global ice coverage \citep{joshi2012suppression, shields2013effect, shields2014spectrum}.

Simulations of its climate suggest that TRAPPIST-1e is one of the most promising candidates for hosting surface liquid water, including across a variety of atmospheric compositions \citep{wolf2017assessing, turbet2018modeling, lincowski2018evolved, sergeev2022trappist} and rotational configurations \citep{chen2023sporadic}. Despite the fact that \cite{greene2023thermal} and \cite{zieba2023thick} found little to no signs of an atmosphere on planets b and c, respectively, there is potential for a secondary atmosphere on planet e \citep{KrissansenTotton, glidden2025jwst}. Notably, planet e may be subject to sporadic tidal interactions from its neighbors that cause millennia-long synodic days \citep{revol2024drifts}. Hence, TRAPPIST-1e is an ideal case study for testing how near-synchronous rotation may impact water-trapping in ice sheets on potentially habitable rocky worlds.

In this study, we present GCM simulations of the spatial distribution and thickness of ice sheets on TRAPPIST-1e, in particular, for the case of a land planet where the global water inventory is limited. We first revisit the case of a synchronously rotating planet to improve estimates of nightside ice thicknesses. We next explore the non-synchronous case. Our work is the first to use a 3D GCM to quantify the water-trapping capabilities of Earth-like land planets around M dwarfs that are subject to near-synchronous rotation. We examine how water inventory and synodic day length influence ice accumulation on these planets. The paper is structured as follows: Section~\ref{methods} enumerates the components of the numerical climate model, Sections~\ref{sec:coupled} and~\ref{sec:chaotic_rot} showcase our ice sheet simulations for synchronous and near-synchronous rotators, Section~\ref{sec:Discussion} speculates on the impact of chaotic spin changes and implications for habitability, and in Section~\ref{sec:conclusions}, we conclude.

\section{Methods} \label{methods}
\begin{table*}[!t]
\centering
\renewcommand{\arraystretch}{1.0}
\caption{Stellar and planetary parameters used in GCM simulations of TRAPPIST-1e.} \label{tab:params}
\begin{tabular}{l c c} 
\toprule
\textbf{Parameter} & \textbf{Value} & \textbf{Units} \\
\midrule
Planet radius* & 5868 & km \\
Gravity* & 8.01 & m s$^{-2}$ \\
\midrule
Orbital period* & 6.1 & Earth days \\
Rotation mode & synchronous, slow rotation & --- \\
Synodic day length & $\infty$, 365, 180, 90 & Earth days \\
Rotation rate & $1.19197 \times 10^{-5}$ & rad s$^{-1}$ \\
Obliquity & 0 & deg \\
Eccentricity & 0 & --- \\
\midrule
Ice albedo & wavelength-dependent & --- \\
Bare rock albedo & 0.2 & --- \\
Surface topography & flat & --- \\
Ground thermal inertia & 500 & J m$^{-2}$ s$^{-1/2}$ K$^{-1}$ \\
H$_{2}$O Global Equivalent Layer & 0, 0.1, 1, 10, 100, 500 & m \\
\midrule
Stellar effective temperature* & 2566 & K \\
Instellation* & 879 & W m$^{-2}$ \\
\bottomrule
\end{tabular}
\vspace{2pt}
\par\noindent
\small{* from \cite{agol2021refining}.}

\end{table*}

\subsection{The LMD Generic Planetary Climate Model}
We performed simulations using the Generic Planetary Climate Model (G-PCM), a 3D GCM developed at the Laboratoire de Métérologie Dynamique. Derived from the LMDz 3D Earth GCM \citep{hourdin2006lmdz4}, the G-PCM is used for modeling paleoclimate and exoplanetary environments. The dynamical core solves the primitive equations of geophysical fluid dynamics on a spherical Arakawa C-grid using a finite-difference method. In addition to simulating the climate of multiple Solar System bodies, including Earth \citep{charnay2013exploring,leconte2013increased,bhatnagar2026fast} and the giant planets \citep[e.g.,][]{bardet2022saturn,boissinot2024jupiter,clement2024storms}, the model has also been used to simulate the paleoclimates of Titan, Venus, and Mars \citep{forget20133d, charnay2014titan, turbet2021day,kuzucan2026early}. These applications have motivated its use for studying a variety of terrestrial worlds around M dwarf stars, including those with a full water-cycle \citep[e.g.,][]{wordsworth2011gliese, leconte20133d, bolmont2016habitability, turbet2018modeling, sergeev2022trappist}. \cite{maurel2025constraints} were the first to simulate a near-synchronous rotational state with the G-PCM to investigate the impact on CO$_{2}$ atmospheric collapse. We extended this framework to examine the impact of slow substellar point drift on ice sheet formation. Water-trapping in ice sheets on synchronously rotating planets has previously been explored with the model \citep{leconte20133d, turbet2016habitability}, but this study is the first to investigate the impact of deviations to the synchronicity.

In this study, we modeled TRAPPIST-1e as an idealized land planet with a flat bare rock surface and an Earth-like atmosphere (1 bar N$_{2}$, 376 ppmv of CO$_{2}$). We set obliquity and eccentricity to zero to remove orbital-parameter-induced variability and to isolate the climatic impact of slow rotation. This is a reasonable approximation given the strong tides raised by the host star in compact planetary systems \citep{heller2011tidal, turbet2018modeling, revol2024drifts}. We used the TRAPPIST-1 emission spectrum computed by \cite{peacock2019predicting}. Table~\ref{tab:params} outlines our physical parameters, which includes stellar and planetary characteristics taken from \cite{agol2021refining}. 

All simulations used a horizontal resolution of 64$\times$48 in longitude$\times$latitude (5.625°$\times$3.75° resolution) and a vertical resolution of 30 layers. The vertical layers are distributed unevenly using hybrid $\sigma$-pressure coordinates (where $\sigma$ is the ratio of atmospheric pressure to surface pressure) that allow for precision at the ground level without compromising the dynamical representation at higher altitudes. The dynamical time-step was 90 model seconds, the physical time-step was 15 minutes, and the radiative timestep was 90 minutes, following past work with the G-PCM \citep[e.g.,][]{leconte20133d, turbet2022trappist, bhatnagar2026fast}. The physical package of the G-PCM computes radiative transfer using the correlated-k method \citep{fu1992correlated, eymet2016kspectrum}. We used the precomputed correlated-k coefficient tables from \cite{leconte2013increased} designed for an Earth-like atmosphere. We also account for sub-grid scale processes following the parameterizations in \cite{leconte2013increased}, including turbulent mixing and moist convection.

\subsection{Water Cycle}
The G-PCM can simulate a water cycle, including transport of water vapor and ice tracers in the atmosphere, surface hydrology, and radiative and thermodynamic effects. Water vapor and ice are treated as variable constituents that can evolve freely between the surface and atmosphere and change phase. The surface liquid and ice reservoirs are partitioned based on the thermodynamic conditions. Condensation to surface ice is assumed to occur when the temperature is below 273 K. Cloud formation and precipitation in our simulations follow the set-up of \cite{charnay2013exploring}, including fixing the radius for the cloud condensation nuclei of liquid water and water ice clouds to 12 $\mu$m and 35 $\mu$m, respectively. For this land planet numerical experiment, we do not simulate a global ocean or sea ice. This is planned for future work using the dynamical slab ocean model \citep{bhatnagar2026fast} developed for the G-PCM.

The G-PCM self-consistently accounts for the wavelength-dependence of albedo by using a simplified spectral albedo law that is calibrated to get ice and snow bolometric albedos of 0.65 around a Sun-like star \citep{joshi2012suppression, bhatnagar2026fast}. Around TRAPPIST-1, the average ice and snow bolometric albedo is 0.25. Surface albedo for these simulations is a combination of the bare rock albedo (here 0.2) and an ice thickness-dependent contribution. Ice accumulation linearly increases the surface albedo of the ground until it reaches a threshold density (fixed to 33 kg m$^{-2}$), above which surface albedo is determined by the ice properties. 

\section{Coupled Atmosphere+Ice Model For A Synchronous Rotator}
\label{sec:coupled}

We first revisit the assessment of nightside ice on a synchronously rotating exoplanet presented in \cite{menou2013water}. We simulated TRAPPIST-1e as a synchronously rotating planet, initialized with a flat topography, a surface albedo of 0.2, and no water cycle. Due to relatively weak heat redistribution and the absence of an atmospheric water vapor greenhouse effect, this idealized synchronous set-up establishes an upper bound on possible ice accumulation \citep{turbet2016habitability}. This configuration also acts as a control numerical experiment to compare with the water-trapping capabilities of the chaotically rotating exoplanets simulated in Sec.~\ref{sec:chaotic_rot}. The simulation was considered converged when the global mean surface temperature and top-of-atmosphere radiative balance stabilized, with temperature varying by 0.04 K and radiative imbalance averaging 0.29 W m$^{-2}$ over the last two Earth years of runtime.

\cite{menou2013water} showed for synchronously rotating exoplanets that nightside ice accumulation is limited at first order by basal melting of the ice sheet, which is determined by surface temperature and geothermal heat flux from the planet's interior. This thermodynamic limit on ice accumulation follows the formulation in \cite{abbot2011steppenwolf}:

{\large
\vspace{-0.5em}
\begin{equation}\label{basalmelteqn}
z_{\rm ice}
=
\frac{A}{F_{\rm geo}}
\ln\!\left(
\frac{
T_{\rm melt}
}{
T_{\rm surf}
}
\right),
\end{equation}
}

\noindent where $z_{\rm ice}$ is the basal-melting-limited ice thickness in meters, $F_{\rm geo}$ is the geothermal heat flux, $T_{\rm melt}$ is the water melting temperature at the ice sheet base in K, $T_{\rm surf}$ is surface temperature in K that is output from the GCM, and $A=651$ Wm$^{-1}$ \citep{petrenko2002physics}. $T_{\rm melt}$ is fixed to 260 K to represent the high-pressure water melting temperature beneath kilometers-thick nightside ice. Geothermal heat flux for exoplanets is empirically unconstrained and theoretically uncertain due to unknown age, internal structure, and composition \citep{turbet2018modeling, barr2018interior, dobos2019tidal, ojha2022liquid, bolmont2026estimation}. Here, geothermal heat flux is assumed to be Earth-like (0.09 W m$^{\rm -2}$ from \citealp{davies2010earth}), which is consistent with the lower bound estimates of the tidal heat flux for planet e \citep{bolmont2026estimation}. Using this equation, we can project global surface temperatures into a steady-state global ice volume.

\begin{figure}[!t]
    \centering
    \includegraphics[width=1\textwidth]{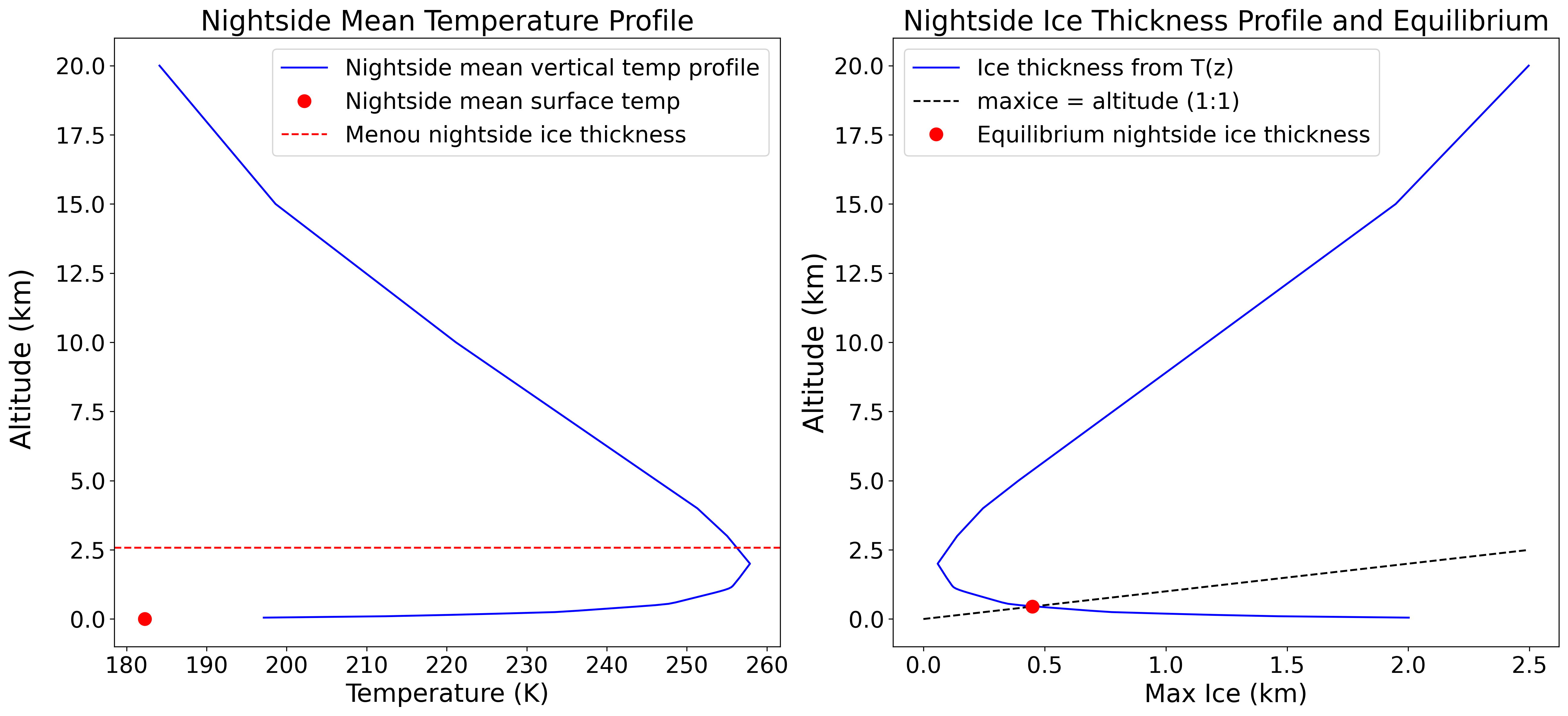}
    \caption{\textit{Left}: Discrepancy in the Menou method when computing the maximum allowable ice thickness. The blue curve is the mean nightside temperature-altitude profile for TRAPPIST-1e as a dry, flat synchronously rotating planet, starting at the first atmospheric level. The red dot indicates the mean nightside surface temperature and the red dashed line is the corresponding mean nightside ice thickness (2.6 km) predicted by the Menou method using Equation~(\ref{basalmelteqn}). \textit{Right}: Maximum allowable ice thickness from the nightside vertical temperature profile. The blue curve is the mean nightside ice thickness-altitude profile computed from the mean nightside temperature-altitude profile. The red dot indicates the intersection where the mean nightside ice thickness matches the altitude level (450 m). Accounting for the nightside vertical temperature inversion substantially decreases the steady-state ice thickness prediction.}
    \label{fig:combined}
\end{figure}

We will call this first-order approximation of ice thicknesses from surface temperatures the Menou method. However, this method neglects two critical feedbacks. Firstly, at higher altitudes, atmospheric temperature and moisture levels change. In Earth’s troposphere, for example, rising air cools and dries as it ascends, limiting precipitation. On a synchronously rotating planet, nightside temperatures increase with altitude \citep{joshi1997simulations, merlis2010atmospheric, labonte2020sensitivity}—a temperature inversion—which could weaken or destabilize this feedback. Figure~\ref{fig:combined} (left) illustrates the discrepancy between surface temperature and the vertical atmospheric temperature structure from the GCM simulation. Following the Menou method with Equation~(\ref{basalmelteqn}), the mean nightside surface temperature of 182~K predicts a mean nightside ice thickness of 2.6 km. However, due to the temperature inversion, the mean nightside temperature at 2.6 km is actually 256~K, implying that this Menou prediction is a considerable overestimation. The vertical temperature profile provides a more physically consistent result: at each altitude, we can map atmospheric temperature into a maximum allowable ice thickness using Equation~(\ref{basalmelteqn}). For this GCM run, the mean nightside vertical temperature profile produces a maximum allowable nightside ice thickness profile as a function of altitude shown in Figure~\ref{fig:combined} (right). The equilibrium mean nightside ice thickness is determined by where the maximum allowable ice thickness equals the altitude, which is only 450 m (Figure~\ref{fig:combined}; right).

An additionally overlooked feedback in the Menou computation is the adjustment of the solid surface beneath the ice load. As ice sheets grow and retreat, the redistribution of mass at the surface causes the mantle and lithosphere to deform and the underlying bedrock to change height. This deformational response of the solid planet, known as glacial isostatic adjustment, is viscoelastic, with an initial smaller and relatively instantaneous elastic response and a slower viscous flow of the planet’s mantle over much longer timescales. On Earth, this viscous adjustment typically takes place over hundreds to thousands of years, though it can be as fast as decades in some regions \citep{peltier1974impulse, barletta2018observed}. Deformation of the surface may continue after tens of thousands of years, but most of it occurs within the first thousand years. Because the mantle viscosity and lithospheric thickness of the planets considered here are unconstrained, we adopt a first-order simplification and assume the ice load is applied slowly enough for the solid surface to be fully adjusted at each timestep. We assume that this surface response is local, neglecting surface deflections away from the point load because they are small relative to the deflection beneath the load. The resulting isostatically-adjusted ice surface elevation, $z_{\rm elev}$, can be approximated to first-order by Airy isostasy \citep{airy1855}:

{\large
\vspace{-0.5em}
\begin{equation}\label{isostasy}
    z_{\rm elev} = \left(\frac{\rho_{\rm mantle}-\rho_{\rm ice}}{\rho_{\rm mantle}}\right)z_{\rm ice},
\end{equation}
}

\noindent where $\rho_{\rm mantle}$ is the density of the mantle (3300 kg m$^{-3}$), $\rho_{\rm ice}$ is the density of ice (1000 kg m$^{-3}$), and $z_{\rm ice}$ is the steady-state ice thickness in meters computed in Equation (\ref{basalmelteqn}). The magnitude of the deformation depends on the density contrast between the mantle and ice. On Earth, glacial isostatic adjustment critically impacts the stability, dynamics, and evolution of ice sheets and their interactions with the atmosphere and climate system (see \citealp{whitehouse2018glacial} for a review). 
In our model, since ice dynamics are not included and accumulation depends only on surface climate and ice surface elevation, this feedback works directly through an ice elevation feedback: as the ice load grows, the land beneath it sinks, lowering the ice surface to a cooler atmospheric level, in turn influencing how the ice sheet evolves.

\begin{figure*}[!t]
    \centering
    \includegraphics[width=\textwidth] {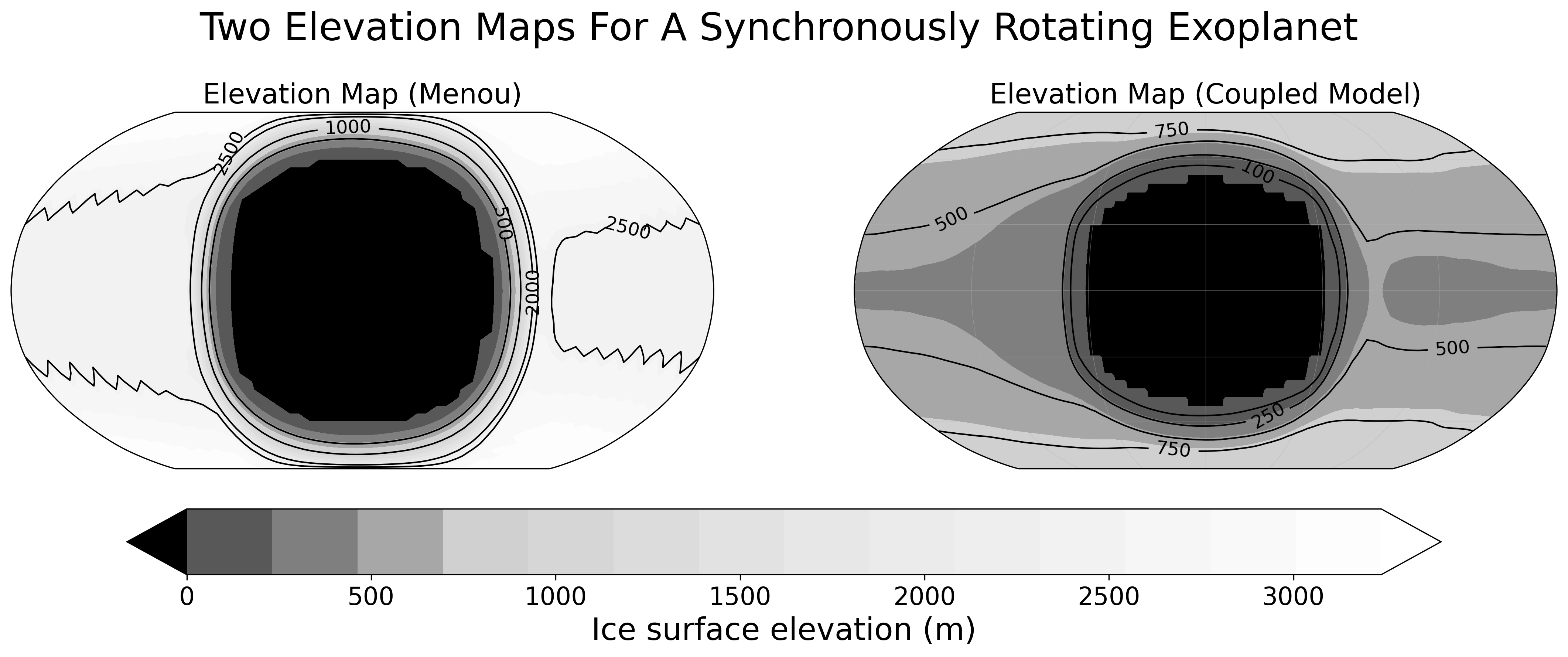}
    \caption{Two elevation maps for a synchronously rotating exoplanet. Black represents flat, ice-free land. \textit{Left}: Ice elevation map of maximal ice sheet growth limited exclusively by basal melting based on GCM surface temperatures following \cite{menou2013water}. \textit{Right}: Surface elevation is solved for iteratively, accounting for the evolving vertical temperature profile and glacial isostatic adjustment. The large difference between elevation maps demonstrates that a coupling between climate and ice models is necessary.}
    \label{fig:elevationcomparison}
\end{figure*}

To provide a more robust assessment of maximum nightside water-trapping, we coupled the G-PCM to a 1D ice model, starting from the converged state of the dry, flat synchronously rotating planet. In each coupling loop the G-PCM was run for one Earth year of simulation time, and the resulting mean vertical temperature profile was input into the ice model. Using Equation (\ref{basalmelteqn}), the ice model computed the altitude at which basal melting occurs. This steady-state ice thickness was then used in Equation (\ref{isostasy}) to compute the surface topography, which was the prescribed surface boundary condition in the following GCM run. We performed this coupling for 24 loops until the ice volume stabilized. The equilibrium ice surface elevation map from the coupled model is shown in Figure~\ref{fig:elevationcomparison} (right). This topographical map exhibits considerably lower ice surface elevation than the Menou map (Fig.~\ref{fig:elevationcomparison}; left) due to thinner ice and accounting for bedrock subsidence in response to the ice load. The Menou method prediction from surface temperatures yields a global ice volume that is $\sim$3 times larger than the coupled model, trapping the equivalent of 55$\%$ of Earth's surface water.

Table~\ref{tab:MenouvsCoupled} shows ice and elevation values computed with increasing model complexity. The first column follows \cite{menou2013water}, and the last column corresponds to the coupled model. The intermediate columns isolate the respective impacts of the atmospheric profile and isostasy. Our results reveal that the final ice sheet elevation depends on the whole tropospheric temperature structure and lithospheric response. The ice volume is expressed in terms of a global equivalent layer (GEL), defined as the average depth of a global water layer if the entire inventory is spread uniformly across the surface as liquid.~\footnote{The GEL of Earth's oceans is 2700 m \citep{charette2010volume}.}

On a synchronously rotating exoplanet, conditions are static due to the time-invariant irradiation pattern. Nightside ice sheets therefore reach a steady-state relatively quickly when running the coupled model. Conversely, for a slowly rotating planet, this coupled model framework is computationally prohibitive due to the long run time required to assess how a long diurnal cycle impacts ice sheet formation and surface deformation, which evolve over millennia. We therefore note that the relative error between the Menou method and the coupled model for the synchronous case can provide a reference for the error when using the Menou method to approximate maximum ice thicknesses on slowly rotating exoplanets (Sec.~\ref{sec:chaotic_rot}).

\begin{table*}[!h]
\centering
\small
\caption{Ice and elevation diagnostics from models with increasing complexity for a synchronously rotating planet.}
\label{tab:MenouvsCoupled}

\begin{tabular*}{\textwidth}{@{\extracolsep{\fill}}|l |rr | rr | r|}
\hline
Quantity & Menou* & Menou + Isostasy & 3D Atm & 3D Atm + Isostasy & Coupled Model* \\
\hline
Thermo-Limit Ice (m GEL) & 1486 & 1486 & 283 & 283 & 540 \\
Frac Earth oceans in ice  & 0.55 & 0.55 & 0.10 & 0.10 & 0.20 \\
Peak ice (m) & 3185 & 3185 & 910 & 910 & 1304 \\
Mean elevation (m) & 1486 & 1206 & 283 & 205 & 376 \\
\hline
\end{tabular*}
\vspace{2pt}
\par\noindent
\small{* are plotted in Fig.~\ref{fig:elevationcomparison}.}
\end{table*}

\section{Ice Sheets On Near-Synchronous Rotators} \label{sec:chaotic_rot}
\subsection{Simulation Set-up}\label{sec:setup}
We explored the relationship between synodic day length and water-trapping in ice sheets for near-synchronously rotating exoplanets. The simulation of synodic day lengths of $>$1 Earth year was computationally prohibitive due to the extended runtime required to capture a single diurnal cycle. Therefore, we imposed three relatively short synodic day lengths to approximate the near-synchronous regime: 3 months (90 days), 6 months (180 days), and 1 year (365 days) in Earth time. Three months was the shortest day length tested for which the daily-mean atmospheric circulation and surface temperatures stabilized close to the climatology of the synchronously rotating planet, exhibiting the climatic regime of a longer period rotational state. We could therefore scale up our analysis of water-trapping for the shorter simulated day lengths to longer day lengths (see Sec.~\ref{sec:longerdaylengths}). Nevertheless, it is worth mentioning that differences in the equatorial daily mean surface temperature profiles for the different day lengths show that this approximation of the near-synchronous rotation regime is imperfect (see Appendix, Fig.~\ref{fig:tsurfvslon}). The larger eastward longitudinal offset in the warmest surface temperatures for shorter synodic days indicates that surface thermal inertia still influences the climate response, unlike in the synchronous case where the surface has infinite time to react to the stellar forcing.

Near-synchronously rotating exoplanets may experience substellar point drift in the prograde (westward) or retrograde (eastward) direction across the surface, depending on if the orbital rate is perturbed slightly slower or faster than the spin rate, respectively \citep{revol2024drifts}. Due to computational expense, for our experiment, we only tested the prograde case as a representative spin regime. We simulated the substellar point completing a full circulation, or synodic day, and did not consider the climatic impact of the substellar point oscillating around 0° longitude known as libration.

We initialized our suite of near-synchronous rotation simulations using the equilibrated state from the flat and dry synchronous run to reduce spin-up time. Subsequently, from the equilibrated state of the dry, near-synchronous rotation runs we prescribed a uniform surface liquid water reservoir. For each day length, we tested five water inventories, reported in terms of a global equivalent layer, which span $\sim$10$^{-5}$--10$^{-1}$ fraction of Earth's inventory (Table~\ref{tab:params}). Each case was essentially a ``swamp'' planet, with a shallow liquid water reservoir globally distributed. To minimize computational expense, after the initial spin-up from the uniform layer, we preserved the spatial distribution of the ice but scaled the ice field by a uniform factor to extrapolate toward the trend in the global ice volume. This extrapolated state was initialized as the new guess and evolved toward equilibrium. We considered all wet simulations to reach quasi-equilibrium when the ice volume varied by fewer than three centimeters between successive synodic days.

\subsection{Simulations of the Near-Synchronous Regime}\label{nearsyncsims}
\begin{figure}[t]
    \centering
    \includegraphics[trim=100 250 100 150, clip, width=\linewidth]{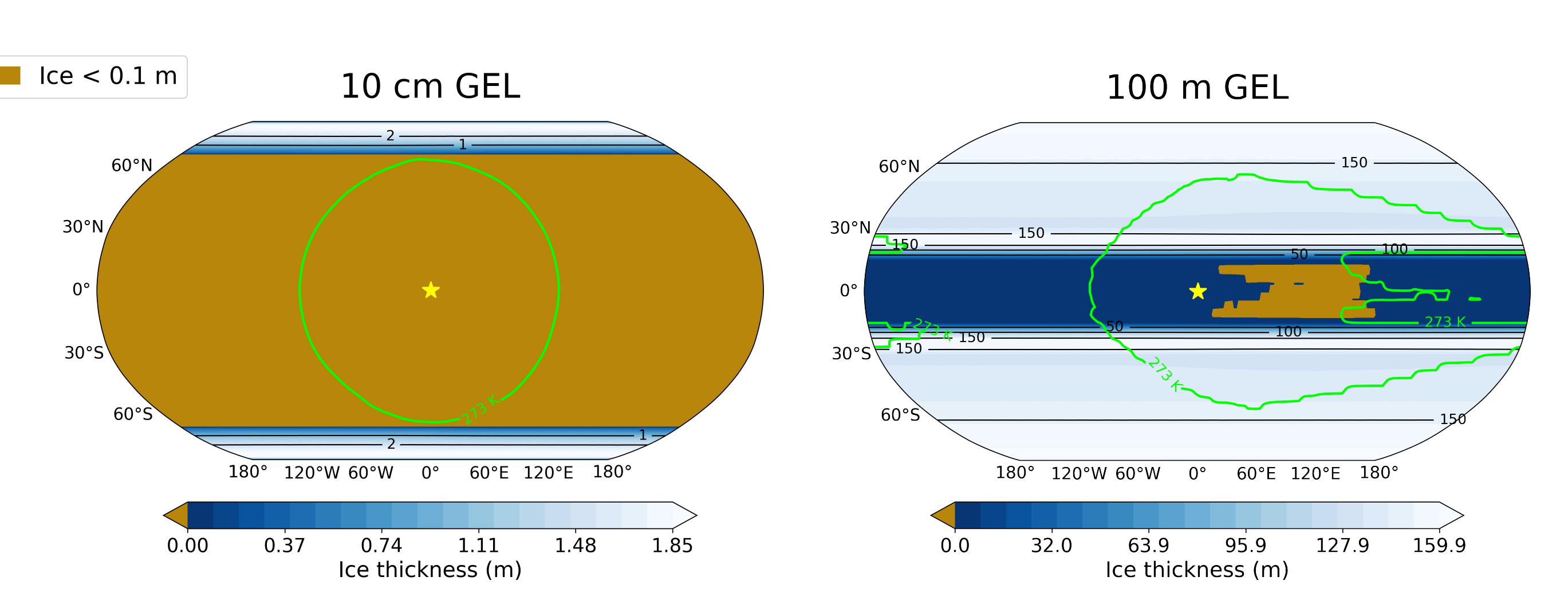}
    \caption{Two example ice maps, assuming a 6-month synodic day, averaged over the final synodic day of the simulation. \textit{Left}: 10 cm GEL water inventory. \textit{Right}: 100 m GEL water inventory. Both maps are in the star-centered frame of reference. Brown represents ice-free land and green is the 273 K temperature contour. The yellow star represents the substellar point.}
    \label{fig:dailyavgice}
\end{figure}

\begin{figure}
    \centering
    \includegraphics[width=0.75\linewidth]{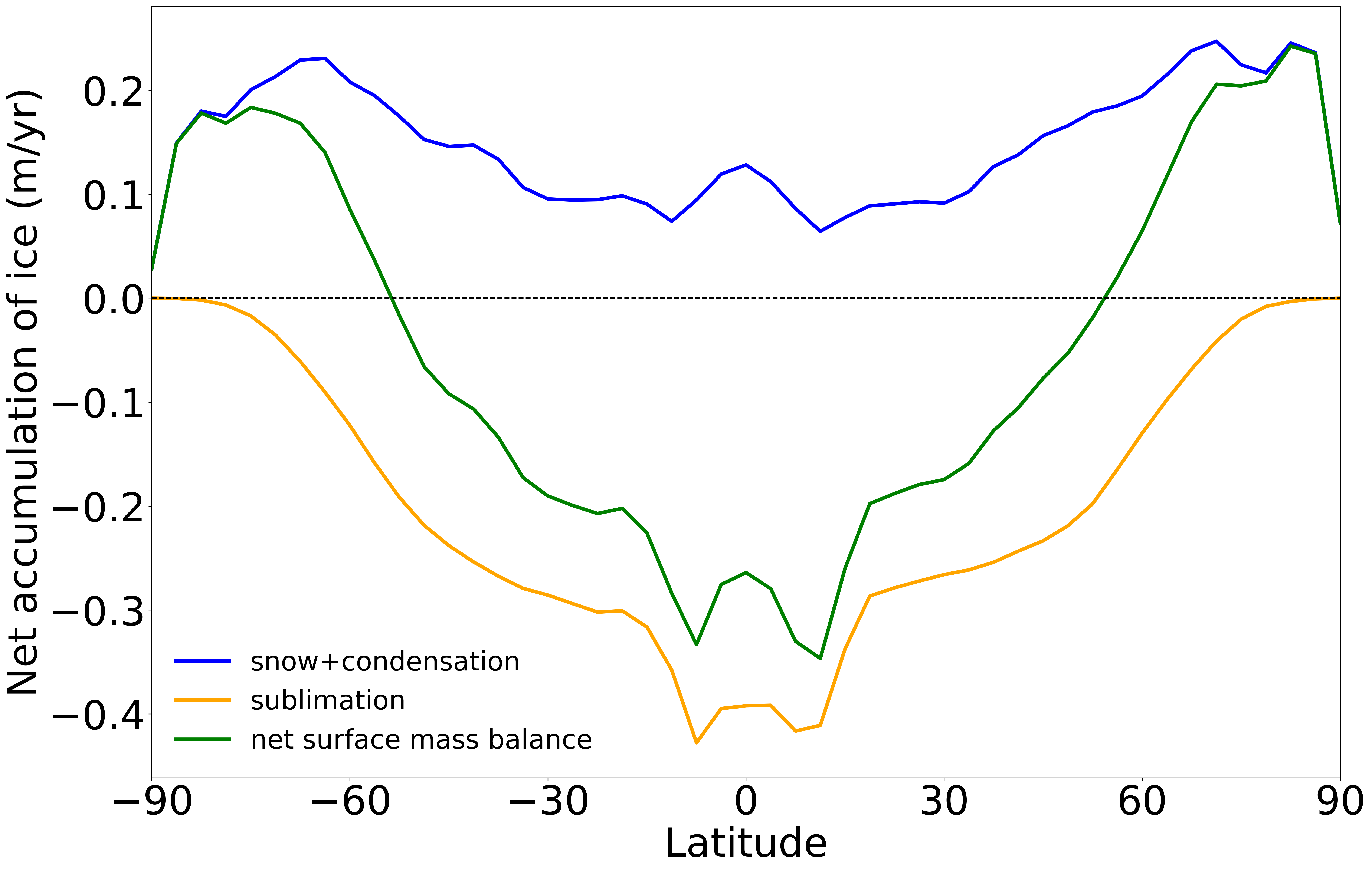}
    \caption{Net surface water mass balance and individual components as a function of latitude for the 6-month synodic day, 100 m GEL case. Each value corresponds to the median time-averaged rate across all longitudes at each latitude, with the time average taken over the final three synodic days of the simulation. Accumulation wins out at latitudes poleward of $\pm$$54^\circ$, while sublimation causes a net loss of ice at lower latitudes.}
    \label{fig:net_accum}
\end{figure}

For the 3-month, 6-month, and 1-year synodic day lengths, ice accumulation is predominantly in polar ice caps across all water inventories. Figure~\ref{fig:dailyavgice} shows two example ice maps computed by the Generic-PCM for TRAPPIST-1e, assuming a 6-month synodic day and a 10 cm GEL water inventory (left panel) and 100 m GEL water inventory (right panel). Both maps are centered on the substellar point and averaged over the final day of the simulation. Though the volume of ice for these two planets has stabilized by the end of the respective simulations, the spatial distribution of this ice, particularly for the 100 m GEL inventory, is still evolving. The net budget of water with respect to latitude shows a global meridional transport of water poleward (Fig.~\ref{fig:net_accum}). There is net accumulation at latitudes above $\pm54^\circ$, while at low latitudes, there is a net loss of ice due to sublimation. While reaching full spatial convergence was too computationally expensive to capture with the G-PCM, we note that the relatively small, eastward offset ice-free region for the 100 m GEL inventory should expand into a broader low-latitude ice-free band with time.

To assess the maximum water-trapping in, and steady-state spatial distribution of, ice across the suite of simulations, we computed the thermodynamic limit. We followed the Menou method, but adapted Equation~\ref{basalmelteqn} to account for the pressure-dependence in the water melting temperature by prescribing,
\vspace{+0.3mm}
\begin{equation}
T_{\mathrm{melt}}
=
13\,\mathrm{K}\,
\exp\left(-\frac{z_{\mathrm{ice}}}{1000\,\mathrm{m}}\right)
+ 260\,\mathrm{K},
\label{eq:tmelt}
\end{equation}
\vspace{-4mm}

\noindent where thin ice ($z_{\rm ice}$ $\ll$  1 km) melts near 273 K, while at larger thicknesses ($z_{\rm ice}$ $\gg$ 1 km), the melting temperature approaches the high-pressure limit of 260 K (see Appendix, Fig.~\ref{fig:basalmeltcurve}). For the 3-month, 6-month, and 1-year day lengths, the diurnal cycle is short enough that the surface experiences the equivalent of a seasonal cycle, with warm (dayside) months and cold (nightside) months. Because ice sheets accumulate and stabilize over much longer timescales than months-long warm and cold periods, we assume for these planets that ice accumulation is governed by the mean climate state. Therefore, we computed the steady-state ice thickness at each location from the diurnally-averaged surface temperatures using Equation~\ref{basalmelteqn} with the parameterization in Equation~\ref{eq:tmelt}. At each latitude, we took the median ice thickness across all longitudes to exclude longitudinal outliers. While diurnally-averaged surface temperatures may be cold enough to support ice formation, net sublimation nonetheless may prevent accumulation. Therefore, the thermodynamic estimate from Equation~\ref{basalmelteqn} is only relevant and possibly realized for locations with mean net accumulation of ice (i.e., positive mass balance).

\begin{figure}
    \centering
    \includegraphics[trim=100 100 100 150, clip, width=\linewidth]{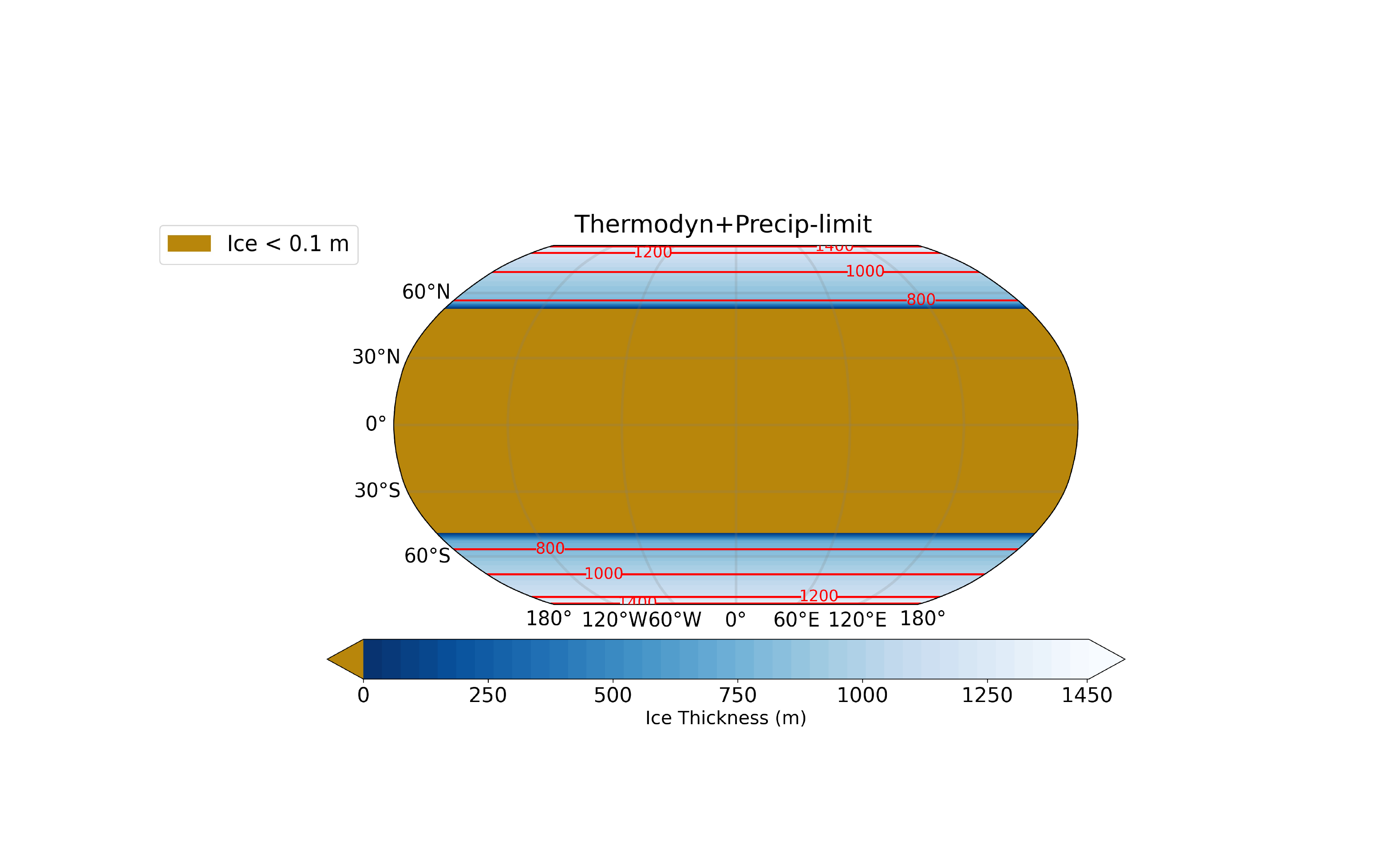}
    \caption{Maximum ice thicknesses constrained by both surface temperature and accumulation for the case of a 6-month synodic day length and 500 m GEL water inventory. The thermodynamically-limited ice thicknesses were computed at all locations in Equation~(\ref{basalmelteqn}), using the surface temperatures averaged over the last three synodic days of the simulation. At each latitude, we took the median ice thickness across all longitudes to reduce the impact of longitudinal outliers. Only latitudes where the median time-averaged net surface mass balance is positive are accounted for. Dark orange shading represents ice-free land. The global equivalent layer of ice is 192 m, and the remainder is available as surface liquid water.}
    \label{fig:maximal6monthice}
\end{figure}

For the four lowest water inventories (0.1–100 m GEL) across each simulated day length, the steady-state predicted global ice volume exceeds the initialized water inventory ($>$ 100 m GEL). Therefore, while basal melting could be a relevant constraint at localized places, the planets are too water-deficient for basal melting to be a relevant geophysical factor on a global scale. Ice formation on such dry planets is therefore inventory-limited rather than thermodynamically-limited. Table~\ref{tab:500mGEL} shows diagnostics for planets with a 3-month, 6-month, and 1-year synodic day length, respectively, for the initialized 500 m GEL inventory. In each case, the maximum possible ice GEL from the net accumulating latitudes is $\sim$150--200 m, indicating that there is sufficient water in the system to accumulate to the maximum limit. Figure~\ref{fig:maximal6monthice} shows the steady-state ice map for a 6-month day length and 500 m GEL water inventory, which demonstrates the expected polar ice distribution.

\begin{table}
\centering
\setlength{\tabcolsep}{4pt}
\renewcommand{\arraystretch}{0.9}
\caption{Ice diagnostics for all day lengths and a 500 m GEL initial water inventory. The ice volume estimates are computed from the median thermodynamically-limited ice thicknesses at each latitude across all longitudes, accounting only for latitudes where the zonal median time-averaged net surface mass balance is positive.} \label{tab:500mGEL}
\small

\begin{tabular}{@{\hspace{0pt}}l r r r@{\hspace{0pt}}}
\toprule
Synodic Day Length & 3 mo & 6 mo & 1 yr \\
\midrule
Thermo+Precip-Limit Ice (m GEL) & 202 & 192 & 172 \\
Frac Earth oceans in ice & 0.06 & 0.06 & 0.05 \\
Peak ice thickness (m) & 1683 & 1454 & 1379 \\
Mean elevation (m) & 141 & 134 & 119 \\
\bottomrule
\end{tabular}
\end{table}

\subsection{Extrapolation to Longer Day Lengths} \label{sec:longerdaylengths}
For day lengths on the scale of hundreds to thousands of Earth years, a planet's surface and climate conditions are increasingly determined by a quasi-permanent dayside and nightside. While simulations of such long day lengths were computationally infeasible, we extrapolated analytically from the shorter period simulations, using the net precipitation and surface temperature profiles. Because ice sheet accumulation becomes governed by prolonged exposure to dayside heating and nightside cooling rather than the diurnally-averaged state, we separated the dayside and nightside hemispheres in our analysis. 

\begin{figure}
    \centering
    \includegraphics[trim=250 175 200 250, clip, width=\linewidth]{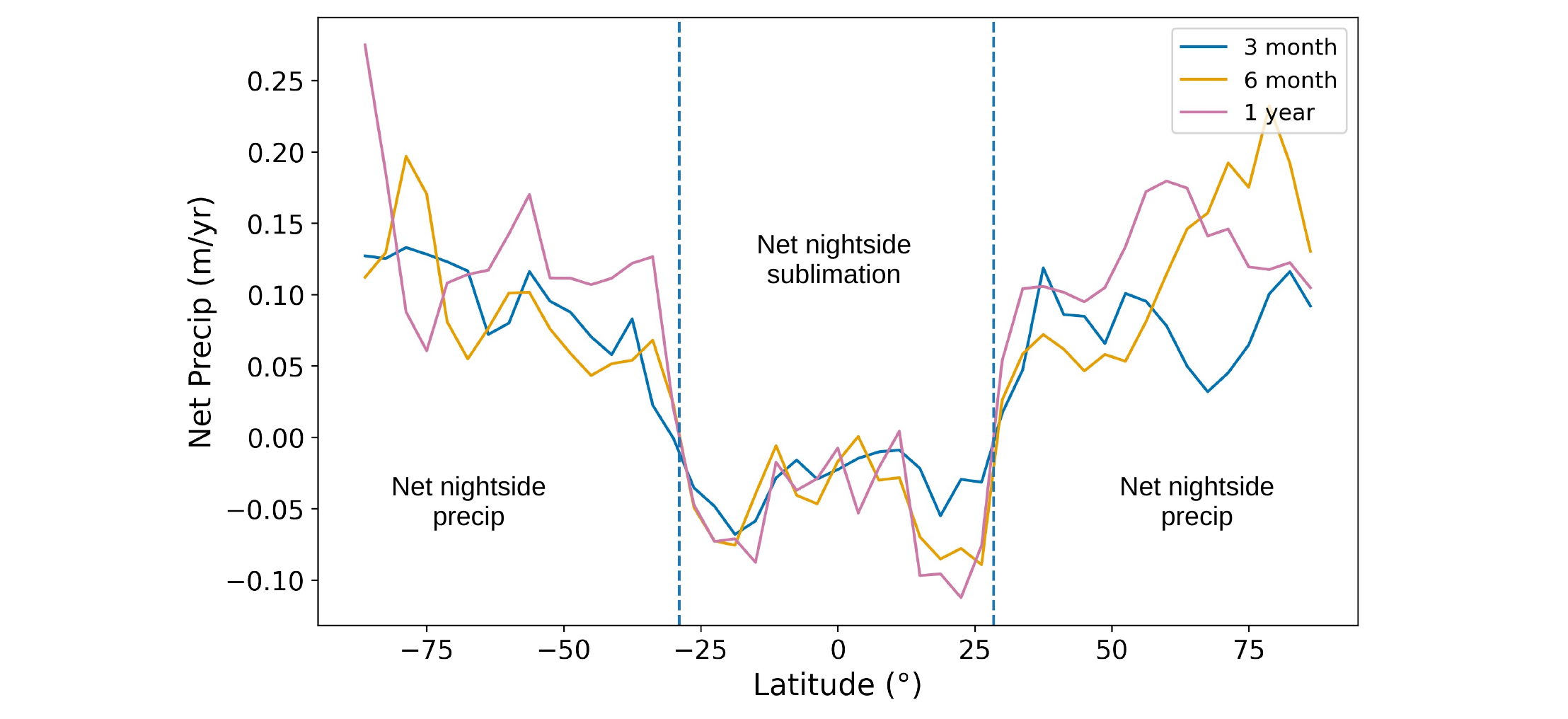}
    \caption{Median nightside net precipitation rates as a function of latitude for the 3-month, 6-month, and 1-year day lengths, assuming a 500 m GEL water inventory. The 1-year day length rates were used to scale up ice thicknesses on the nightside for longer day lengths. The dashed lines represent where the 1-year net precipitation rates go to zero, demarcating the bounds of net sublimation at low latitudes.}
    \label{fig:netprecipnightside}
\end{figure}

Figure~\ref{fig:netprecipnightside} shows the nightside latitudinal profile of net precipitation rates for the 3-month, 6-month, and 1-year day lengths, assuming a 500 m GEL initial water inventory. At each latitude, we calculated the median nightside time mean rate across all nightside longitudes. The net precipitation rates are similar for all day lengths, indicating that cumulative ice accumulation should increase linearly as the nightside duration increases. In other words, the amount of ice that can build up is proportional to the duration of the night. Therefore, using the net precipitation rates from the 1-year synodic day simulation, we could obtain a first-order, precipitation-limited approximation of ice accumulation for any prescribed day length. This ice accumulation is also limited thermodynamically. Using the nightside time mean surface temperatures, we computed ice thicknesses at each location using Equations~(\ref{basalmelteqn}) and~(\ref{eq:tmelt}). We then computed the median nightside ice thickness across all longitudes at each latitude. We took the minimum thickness values between the net precipitation and thermodynamic limits to be representative of the maximum amount of ice that could build up in the extrapolated amount of time (see Appendix, Fig.~\ref{fig:extrapolation}). For each extrapolated day length, we performed this calculation for both the nightside and dayside and summed the respective ice reservoirs together. 

\begin{figure}
    \centering
    \includegraphics[trim=100 270 375 175, clip, width=\linewidth]{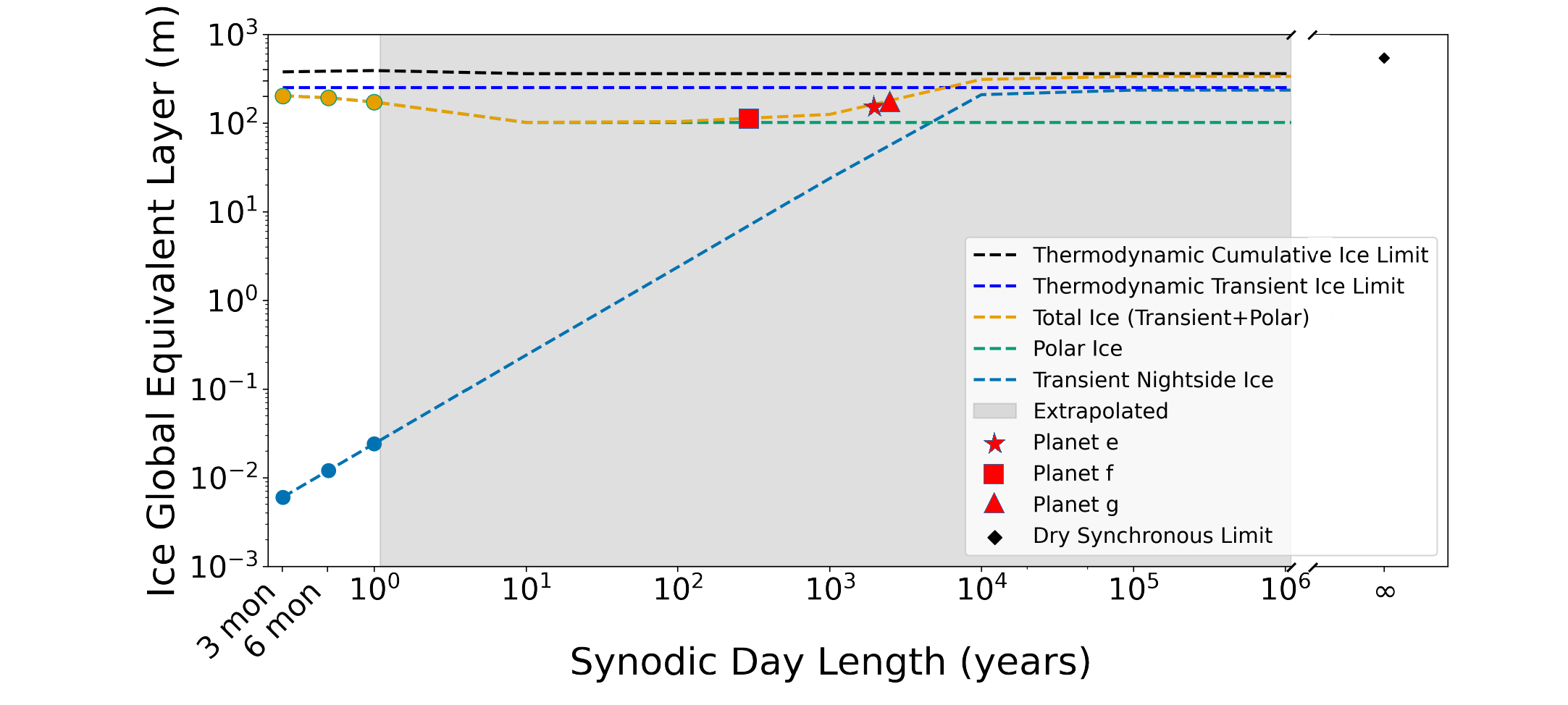}
    \caption{Ice sheet accumulation on TRAPPIST-1e (in terms of a global equivalent layer) as a function of day length, assuming an Earth-like land planet and a 500 m GEL initialized water inventory. Black dashed line: the thermodynamic limit for the total ice volume; dark blue: the thermodynamic limit for the transient ice reservoir; yellow: the sum of all accumulated ice; light blue: interpolated trend in the volume of transient ice on the nightside; green: interpolated trend in the volume of the permanent ice. The left-most data points were computed in Section~\ref{nearsyncsims}, using the diurnally-averaged surface temperature and surface mass balance fields. These represent the short-period end of the near-synchronous regime. The gray region contains the extrapolated data. The total ice is slightly beneath the thermodynamic limit because net sublimation limits snowfall in certain cold regions. The right-most black diamond represents the dry, synchronous case computed using the coupled model in Section~\ref{sec:coupled}. The red shapes correspond to the ice GEL for the habitable zone TRAPPIST-1 planets given example estimated day lengths from \cite{revol2024drifts}.} \label{fig:iceaccumvsdaylength}
\end{figure}

The trend in ice volume across day length and the partitioning between polar (permanent) and transient ice is shown in Figure~\ref{fig:iceaccumvsdaylength}. The left-most data points represent the values from Table~\ref{tab:500mGEL}, computed for the simulated day lengths from the diurnally-averaged surface temperature and mass balance fields. The shaded region corresponds to the extrapolated data.

As TRAPPIST-1e approaches synchronous rotation, the increasingly long nightside exposure enables thicker ice formation at lower latitudes. We refer to this ice as transient because it forms outside of the permanent polar reservoir and sublimates entirely on the dayside. The light blue dashed line in Figure~\ref{fig:iceaccumvsdaylength} shows this upward trend in transient nightside ice formation. While only marginal amounts of transient ice ($\sim$1 cm GEL) form for the shorter period planets, the transient ice volume reaches 208 m GEL for a day length of ten thousand years. Despite the monotonic increase in transient ice, its accumulation is ultimately limited by net sublimation at lower latitudes on the nightside (Fig.~\ref{fig:netprecipnightside}; Fig.~\ref{fig:dailyavgprecipmap}). Conditions at these low latitudes, though cold, are too dry to build up ice, so even with increasingly long nightside exposure, the transient reservoir never fully realizes its thermodynamic limit (dark blue dashed line in Fig.~\ref{fig:iceaccumvsdaylength}). On the dayside, this transient nightside ice reservoir sublimates away to form a liquid water reservoir. The estimated volume of this dayside liquid water reservoir ranges from approximately a fourth to a half of the Atlantic ocean across the different day lengths. However, this volume range likely represents an overestimation either because the simulations have not completely converged or because the initial water inventory (500 m GEL) exceeds the maximum amount of water that can be trapped in ice.

The trend in polar ice accumulation, shown in green, decreases across the simulated day lengths and converges to a common lower ice volume for the extrapolated day lengths. The extrapolated day lengths have a common steady-state polar ice volume because the spatial extent of the permanent reservoir and the steady-state limit on ice accumulation are derived from the 1-year case. The decreasing trend is, in part, due to the reduced latitudinal extent of ice. For the 3-month day, the polar ice reservoir extends down to $\pm$$51^\circ$, for the 1-year day, the polar ice reservoir extends down to $\pm$$55^\circ$, and for the extrapolated day lengths, this reservoir extends to only -$56^\circ$ in the Southern hemisphere and $64^\circ$ in the Northern hemisphere. For the extrapolated day lengths, the permitted polar ice caps are also thinner on the dayside due to relatively higher temperatures from the prolonged stellar irradiation. Therefore, although the peak in ice thickness is relatively uniform across day length ($>$1 km), the combination of a reduced latitudinal extent of polar ice and weakened dayside accumulation ultimately makes the volume of ice caps smaller for the longer synodic days.

\section{Discussion}\label{sec:Discussion}
\subsection{Reaching Equilibrium} \label{sec:timescales}
An important consideration is that near-synchronously rotating exoplanets around M dwarfs may not exhibit only one spin regime through time \citep{vinson2019chaotic, chen2023sporadic, shakespeare2023day, revol2024drifts}, which could prevent them from reaching a steady-state ice volume and spatial distribution. Due to sporadic perturbations from neighboring planets, the substellar point may drift in a new direction or speed within 10$^{1}$--10$^{5}$ years~\citep{vinson2019chaotic, shakespeare2023day}. As shown for the case of prograde rotation, transitions between day lengths could dramatically alter global water-trapping and surface liquid water availability by changing transient ice accumulation by orders of magnitude (Fig.~\ref{fig:iceaccumvsdaylength}). 

Additionally, changes between prograde, retrograde, and libration states could further complicate the possibility for an equilibrium. Within the near-synchronous regime, a retrograde rotating planet should have the same climatology as the prograde case. However, the movement of the surface in the opposite direction relative to the atmospheric circulation could impact where moisture and heat get advected relative to the ice, thereby impacting ice accumulation. Moreover, a transition from a circulating to librating state would bring the planet into a configuration more similar to that of a synchronous rotator, where ice accumulation would remain predominantly permanently on the nightside with sublimation only near the terminators. Exploring the nature of ice formation in these other near-synchronous spin regimes should be a focus of future work.

As planets switch between these different spin behaviors, the substellar point drift may change more rapidly than the timescale of ice accumulation, causing the polar and transient ice reservoirs to perpetually fluctuate and lag behind the evolving steady-state. For example, given the calculated net snowfall rates, the area-weighted accumulation timescale to attain the maximum possible ice at the poles is $\sim$10$^{4}$ years. This means that a steady-state could be reached if a given spin regime lasts for at least $\sim$10$^{4}$ years. Another relevant timescale is the relaxation time of the solid planet towards isostatic equilibrium following ice loading, which can occur over hundreds to millions of years for an Earth-like planet \citep{peltier1974impulse, Ivins2023}.
If the relaxation timescale of the solid planet is slower than the timescale of ice mass changes, the surface topography would likewise perpetually lag behind changes in spin regime and potentially never catch up to the evolving steady-state ice distribution. The equilibrium surface topography and whether it is attainable during a single spin regime will be determined by the planet’s rheological properties as well as the geometry of the ice loading changes. For example, TRAPPIST-1e may have a higher geothermal heat flux than Earth (see \citealp{bolmont2026estimation}), suggesting a less viscous mantle, which would allow for a more rapid response time to an ice load. Table~\ref{tab:500mGEL} shows the mean surface elevation for the 3-month, 6-month, and 1-year synodic day cases computed from Equation~(\ref{isostasy}) assuming that the steady-state ice distribution (\textit{e.g.}, Fig.~\ref{fig:maximal6monthice}) has been reached.

\subsection{Habitability}
\label{sec:habitability}
Though near-synchronously rotating exoplanets may be capable of sustaining substantial dayside surface liquid water, it would not be in the form of stationary lakes or oceans. Rather, to remain on the warm dayside, the liquid water must migrate across the surface as the planet rotates. The prolonged episodes of nightside freeze and dayside thaw are not an inherent obstacle to the origin and evolution of life. 
For example, chemical reactions driven by freeze-thaw cycles on Earth may have caused the first self-replication of RNA, facilitating the development of increasingly complex biomolecules that were needed in the origin of simple cellular life \citep{mutschler2015freeze, zhang2022freeze, attwater2025trinucleotide}. Moreover, the surface of a near-synchronously rotating world may parallel Earth’s permafrost terrain, where the upper surface layer experiences a seasonal freeze and thaw cycle \citep{bonnaventure2013active, dobinski2020permafrost}. On Earth, within this so-called “active layer,” extremophile microbes can enter into a temporary metabolically dormant state to withstand the prolonged freezing and desiccation during cold months \citep{rebecchi2007anhydrobiosis, goordial2016nearing, mackelprang2017microbial}. A near-synchronously rotating exoplanet may ultimately provide a more stable, long-term habitat if organisms can remain within the slowly migrating pool. This would require a globally flat terrain, such that no topographical barriers (\textit{e.g.}, mountains, deep basins) would impede the relocation. However, given that ice formation may change through time depending on the spin regime, the abundance of liquid water on the dayside may ebb and flow. Therefore, even if marine life could undertake this migration, the dayside reservoir may not be as stable a habitat as the oceans on Earth.

\cite{leconte20133d} and \cite{ojha2022liquid} showed that sub-glacial liquid water reservoirs should be present beneath nightside ice sheets on synchronously rotating planets and may be as expansive as the sub-surface oceans on Jupiter’s and Saturn’s icy moons. Depending on a planet's interior composition, radiogenic heating could be sufficient to sustain these vast liquid reservoirs on geological ($>$1 Gyr) timescales, long enough for biological evolution \citep{ojha2022liquid}. On a near-synchronously rotating exoplanet, however, the possibility for sub-glacial reservoirs as a habitable niche is less certain: the basal-melting-constrained steady-state may be short-lived, if reached at all.

\section{Conclusions and Looking Ahead} \label{sec:conclusions}
In this study, we explored the water-trapping capabilities of ice sheets on TRAPPIST-1e, under the assumption of synchronous and near-synchronous rotation. Using a 3D GCM, we demonstrated that ice accumulation on near-synchronously rotating exoplanets is likely not a threat to the availability of surface liquid water even for a relatively dry planet. Even for a synchronously rotating exoplanet, water-trapping in ice is less worrisome than previously thought, as estimations following the methodology of \cite{menou2013water} should be cut by a factor of $\sim$3, from a possible ice global equivalent layer on TRAPPIST-1e of 1486 m to 540 m. 

We found that the global volume and spatial distribution of ice sheets vary with synodic day length, as more transient nightside ice forms at longer day lengths. We note that these ice projections are likely an overestimate, since we did not account for the atmospheric temperature structure or iterative feedbacks with the solid surface, as shown in Sec.~\ref{sec:coupled}. 

Future work should explore a broader parameter space, including more rotational regimes (retrograde, libration), water inventories, atmospheric compositions, surface conditions, and ice dynamics. The presence of an ocean, for example, could limit nightside water-trapping through increased heat transport \citep{hu2014role, del2019habitable}. This analysis is planned for future work using the dynamical slab ocean model \citep{bhatnagar2026fast} developed for the Generic-PCM. Additionally, the gravity-driven flow of ice would have important implications for surface liquid water availability \citep{leconte20133d}. Kilometer-thick polar ice caps can flow over distances of hundreds of kilometers \citep{rignot2011ice}, which could bring ice to the lower latitudes and cause melting at their edges, potentially expanding the dayside liquid water reservoir. 

Ultimately, the long and potentially competing timescales (from decades to multi-millennia) of the diurnal cycle, ice accumulation, surface deformation, and sporadic spin changes make these planets uniquely challenging to simulate with GCMs. Future studies with more robust numerical modeling \citep[e.g.,][]{Clement2024ICM} will be needed to determine the long-timescale evolution of, and feedbacks between, the atmosphere, ice sheets, and solid surface on near-synchronously rotating exoplanets. 

\section{Acknowledgments}
NBC acknowledges support from an NSERC Discovery Grant and a Tier 2 Canada Research Chair. This work was supported by a grant (DOI:~\href{https://doi.org/10.69777/377645}{10.69777/377645}) from the Fonds de recherche du Québec and by the Center for research in astrophysics of Québec (AstroQuébec). The authors also thank the Trottier Space Institute and l’Institut de recherche sur les exoplanètes for their financial support and dynamic intellectual environment. S.S. acknowledges support from the FRQNT Masters Scholarship (DOI:~\href{https://doi.org/10.69777/365826}{10.69777/36582}). S.S. also thanks Ehouarn Millour, Alice Maurel, Robert Fajber, and Colin Goldblatt for insightful conversations during the writing of this work.

\section{Appendix}
\label{app:basalmelt}
\setcounter{figure}{0}
\renewcommand{\thefigure}{A.\arabic{figure}}

\begin{figure}[H]
    \centering
    \includegraphics[width=1\linewidth]{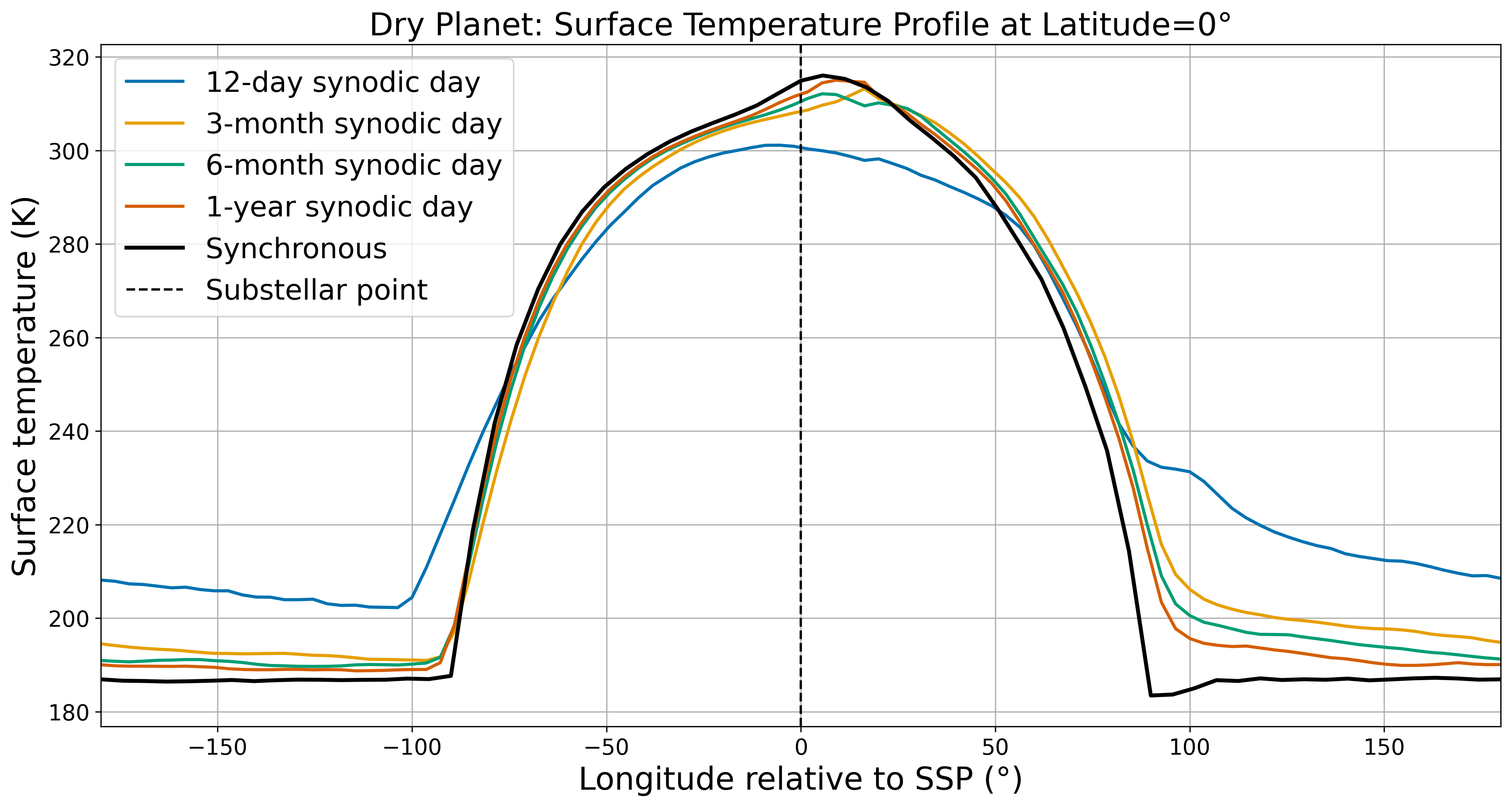}
    \caption{Daily mean surface temperature at the equator (latitude=0°) as a function of longitude for a synchronously rotating planet and slowly rotating planets with synodic day lengths of 3 months, 6 months, and 1 year (Earth time). All slowly rotating planets experience prograde (westward) drift of the substellar point. A planet with a 12-day synodic day length case is also included to represent a rapidly rotating regime that is far from the near-synchronous limit. The longer the synodic day length, the closer the equatorial surface temperature profile approximates the synchronous case (\textit{i.e.}, the impact of surface thermal inertia diminishes as expected).}
    \label{fig:tsurfvslon}
\end{figure}

\begin{figure}
    \includegraphics[width=1\linewidth]{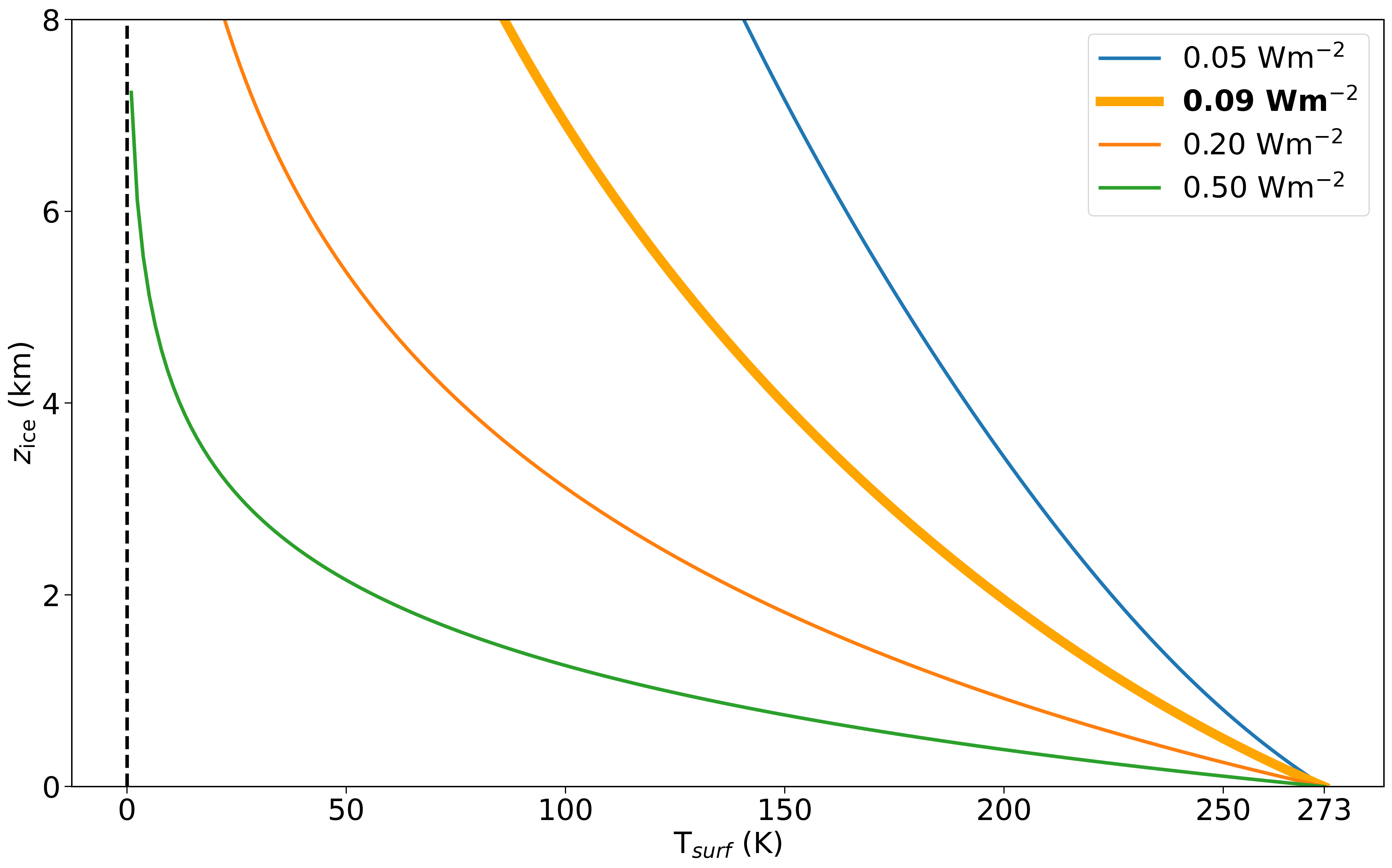}
    \caption{Mapping surface temperature to thermodynamically-limited ice thickness using the pressure-dependent melting temperature parameterization ($\sim$273 K for thin ice; $\sim$260~K for thick ice). Each curve represents the mapping for increasingly high geothermal heat fluxes. The flux value of 0.09 W m$^{-2}$ plotted in bold is the value used in this work.}
    \label{fig:basalmeltcurve}
\end{figure}

\begin{figure}
    \centering
    \includegraphics[width=\linewidth]{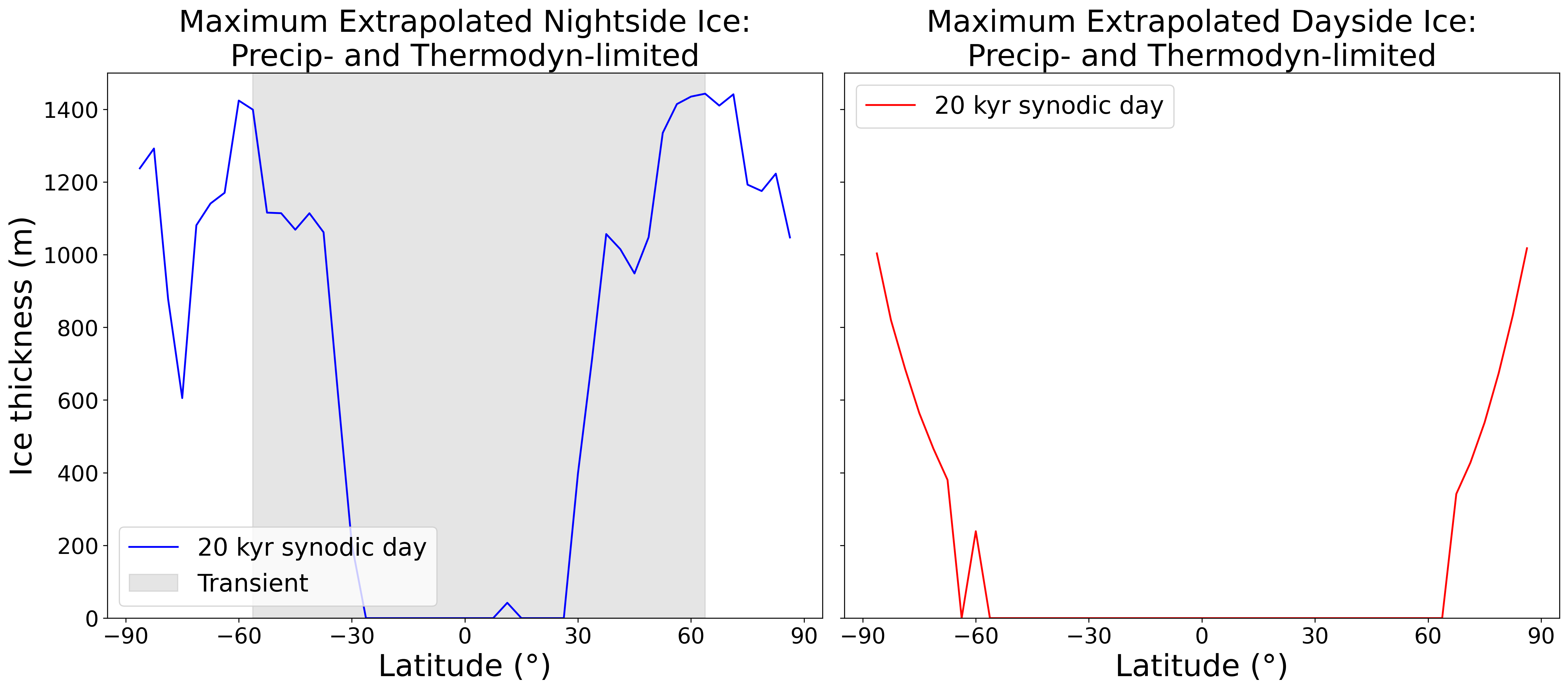}
    \caption{Maximum extrapolated ice thicknesses as a function of latitude, limited by net precipitation and surface temperature. \textit{Left}: Median nightside ice thicknesses across all nightside longitudes at each latitude. The shaded gray region corresponds to the nightside latitudes with transient ice, which sublimates to zero on the dayside. Ice at latitude -63.75° is also transient and goes to zero ice on the dayside. \textit{Right}: Median dayside ice thickness across all dayside longitudes at each latitude. The dayside polar reservoir has less ice than the nightside due to sublimation.}
    \label{fig:extrapolation}
\end{figure}

\begin{figure}
    \includegraphics[trim=200 150 250 190, clip, width=\linewidth]{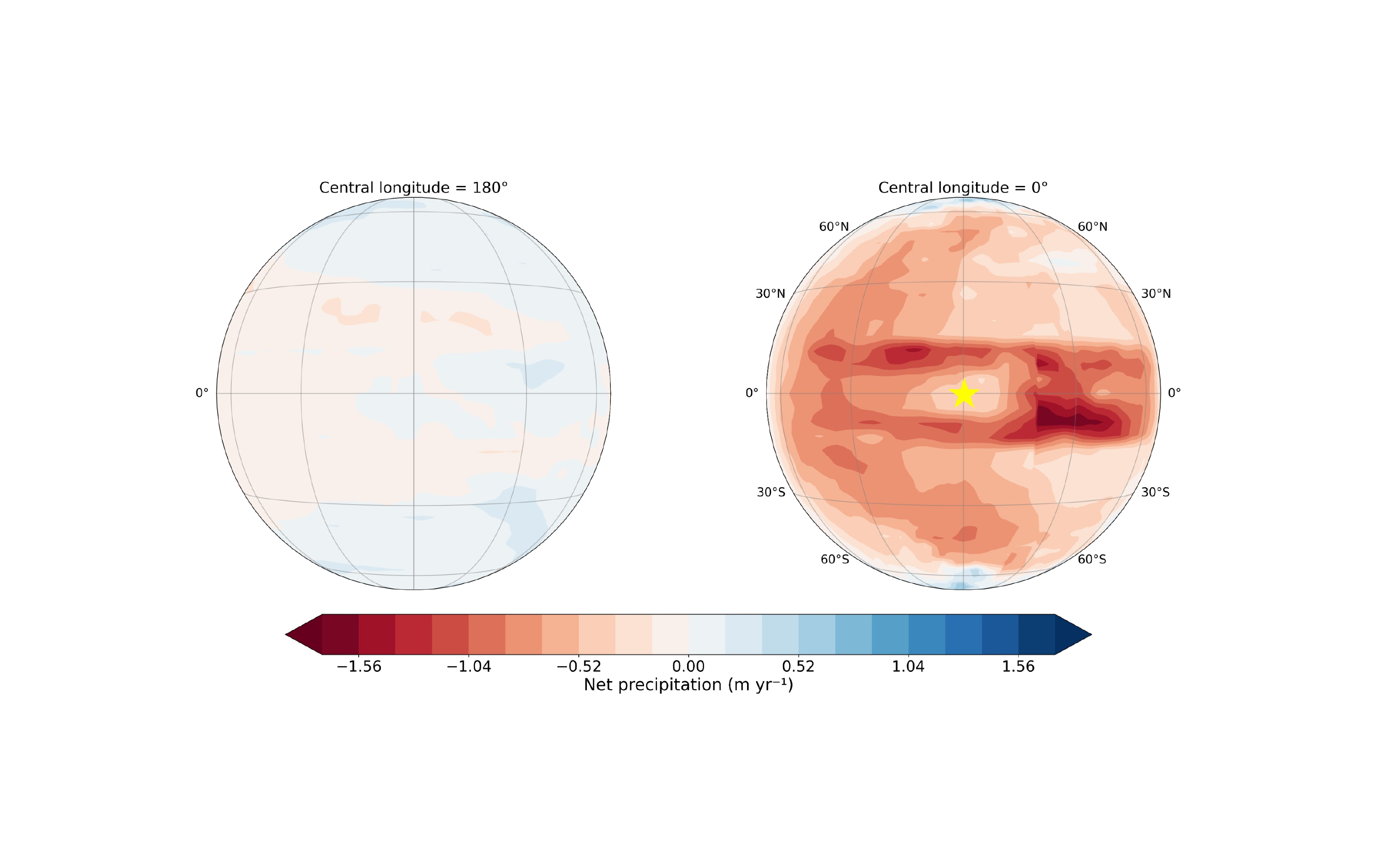}
    \caption{Map of surface mass balance in the star-centered frame, averaged over the final synodic day of the simulation, for the 1-year synodic day, 500 m GEL case. \textit{Left}: Nightside. \textit{Right}: Dayside. The yellow star represents the substellar point.}
    \label{fig:dailyavgprecipmap}
\end{figure}

\FloatBarrier
\bibliography{references}

@ARTICLE{kuzucan2026early,
       author = {{Kuzucan}, Asena and {Bolmont}, Emeline and {Zeyen}, Nina and {Gu}, Wenyu and {Xue}, Jie and {Turbet}, Martin and {Caricchi}, Luca and {Martignier}, Agathe and {Bhatnagar}, Siddharth},
        title = "{Early mars habitability revisited: Oceans, climate dynamics, and microbial survival}",
      journal = {\icarus},
         year = 2026,
        month = nov,
       volume = {459},
          eid = {117236},
        pages = {117236},
          doi = {10.1016/j.icarus.2026.117236},
       adsurl = {https://ui.adsabs.harvard.edu/abs/2026Icar..45917236K}
}

@article{dressing2013occurrence,
  title={The occurrence rate of small planets around small stars},
  author={Dressing, Courtney D and Charbonneau, David},
  journal={The Astrophysical Journal},
  volume={767},
  number={1},
  pages={95},
  year={2013},
  publisher={The American Astronomical Society},
  doi={10.1088/0004-637X/767/1/95}
}

@article{dressing2015occurrence,
  title={The occurrence of potentially habitable planets orbiting M dwarfs estimated from the full Kepler dataset and an empirical measurement of the detection sensitivity},
  author={Dressing, Courtney D and Charbonneau, David},
  journal={The Astrophysical Journal},
  volume={807},
  number={1},
  pages={45},
  year={2015},
  publisher={The American Astronomical Society},
  doi={10.1088/0004-637X/807/1/45}
}

@article{Ivins2023,
  author  = {Ivins, E. R. and van der Wal, W. and Wiens, D. A. and Lloyd, A. J. and Caron, L.},
  title   = {},
  journal = {Geological Society, London, Memoirs},
  year    = {2023},
  doi     = {10.1144/m56-2020-19}
}

@article{peltier1974impulse,
  title={The impulse response of a Maxwell Earth},
  author={Peltier, WR},
  journal={Reviews of Geophysics},
  volume={12},
  number={4},
  pages={649--669},
  year={1974},
  publisher={Wiley Online Library},
  doi={10.1029/RG012i004p00649}
}

@article{KrissansenTotton,
    author = {{Krissansen-Totton}, Joshua},
    title = "{Implications of Atmospheric Nondetections for TRAPPIST-1 Inner Planets on Atmospheric Retention Prospects for Outer Planets}",
    journal = {\apjl},
    year = 2023,
    volume = 951,
    number = 2,
    eid = {L28},
    doi = {10.3847/2041-8213/acdc26}
}

@article{glidden2025jwst,
  author  = {Glidden, A. and Ranjan, S. and Seager, S. and Espinoza, N. and MacDonald, R. J. and Allen, N. H. and Cañas, C. I. and Grant, D. and Gressier, A. and Stevenson, K. B. and Batalha, N. E. and Lewis, N. K. and Long, D. and Wakeford, H. R. and Alderson, L. and Challener, R. C. and Colón, K. and Huang, J. and Lin, Z. and Louie, D. R. and Mullens, E. and Sotzen, K. S. and Valenti, J. A. and Valentine, D. and Clampin, M. and Mountain, C. M. and Perrin, M. and van der Marel, R. P.},
  title   = {JWST-TST DREAMS: Secondary Atmosphere Constraints for the Habitable Zone Planet TRAPPIST-1 e},
  journal = {\apjl},
  volume  = {990},
  number  = {2},
  pages   = {L53},
  year    = {2025},
  doi     = {10.3847/2041-8213/adf62e}
}

@article{greene2023thermal,
  author  = {Greene, T. P. and Bell, T. J. and Ducrot, E. and others},
  title   = {Thermal Emission from the Earth-sized Exoplanet TRAPPIST-1 b Using JWST},
  journal = {\nat},
  volume  = {618},
  pages   = {39--42},
  year    = {2023},
  doi     = {10.1038/s41586-023-05951-7}
}

@article{zieba2023thick,
  author  = {Zieba, S. and Kreidberg, L. and Ducrot, E. and others},
  title   = {No Thick Carbon Dioxide Atmosphere on the Rocky Exoplanet TRAPPIST-1 c},
  journal = {\nat},
  volume  = {620},
  pages   = {746--749},
  year    = {2023},
  doi     = {10.1038/s41586-023-06232-z}
}

@article{kreidberg2025first,
  title={A first look at rocky exoplanets with JWST},
  author={Kreidberg, Laura and Stevenson, Kevin B},
  journal={Proceedings of the National Academy of Sciences},
  volume={122},
  number={39},
  pages={e2416190122},
  year={2025},
  publisher={National Academy of Sciences},
  doi={10.1073/pnas.2416190122}
}

@article{mackelprang2017microbial,
  title={Microbial survival strategies in ancient permafrost: insights from metagenomics},
  author={Mackelprang, Rachel and Burkert, Alexander and Haw, Monica and Mahendrarajah, Tara and Conaway, Christopher H and Douglas, Thomas A and Waldrop, Mark P},
  journal={The ISME journal},
  volume={11},
  number={10},
  pages={2305--2318},
  year={2017},
  publisher={Oxford University Press},
  doi={10.1038/ismej.2017.93}
}

@article{rebecchi2007anhydrobiosis,
  title={Anhydrobiosis: the extreme limit of desiccation tolerance},
  author={Rebecchi, Lorena and Altiero, Tiziana and Guidetti, Roberto},
  journal={Invertebrate Survival Journal},
  volume={4},
  number={2},
  pages={65--81},
  year={2007},
  url={https://api.semanticscholar.org/CorpusID:43921178}
}

@article{goordial2016nearing,
  title={Nearing the cold-arid limits of microbial life in permafrost of an upper dry valley, Antarctica},
  author={Goordial, Jacqueline and Davila, Alfonso and Lacelle, Denis and Pollard, Wayne and Marinova, Margarita M and Greer, Charles W and DiRuggiero, Jocelyn and McKay, Christopher P and Whyte, Lyle G},
  journal={The ISME journal},
  volume={10},
  number={7},
  pages={1613--1624},
  year={2016},
  publisher={Nature Publishing Group},
  doi={10.1038/ismej.2015.239}
}

@article{dobinski2020permafrost,
  title={Permafrost active layer},
  author={Dobi{\'n}ski, Wojciech},
  journal={Earth-Science Reviews},
  volume={208},
  pages={103301},
  year={2020},
  publisher={Elsevier},
  doi={10.1016/j.earscirev.2020.103301}
}

@article{bonnaventure2013active,
  title={The active layer: A conceptual review of monitoring, modelling techniques and changes in a warming climate},
  author={Bonnaventure, Philip P and Lamoureux, Scott F},
  journal={Progress in Physical Geography},
  volume={37},
  number={3},
  pages={352--376},
  year={2013},
  publisher={SAGE Publications Sage UK: London, England},
  doi={10.1177/0309133313478314}
}

@article{rignot2011ice,
  title={Ice flow of the Antarctic ice sheet},
  author={Rignot, Eric and Mouginot, Jeremie and Scheuchl, Bernd},
  journal={Science},
  volume={333},
  number={6048},
  pages={1427--1430},
  year={2011},
  publisher={American Association for the Advancement of Science},
  doi={10.1126/science.1208336}
}

@article{tuomi2019frequency,
  author = {Tuomi, Mikko and Jones, Hugh R. A. and Butler, R. Paul and Arriagada, Pamela and Vogt, Steven S. and Burt, Jennifer and Laughlin, Gregory and Holden, Brad and Shectman, Stephen A. and Crane, Jeffrey D. and Thompson, Ian and Keiser, Sarah and Jenkins, Jon M. and Berdiñas, Z. M. and Diaz, Matias and Kiraga, Marcin and Barnes, John R.},
  title = {Frequency of planets orbiting M dwarfs in the Solar neighbourhood},
  journal={arXiv preprint arXiv:1906.04644},
  year = {2019},
  eprint = {1906.04644},
  doi = {10.48550/arXiv.1906.04644}
}

@article{menou2013water,
  title={Water-trapped worlds},
  author={Menou, Kristen},
  journal={The Astrophysical Journal},
  volume={774},
  number={1},
  pages={51},
  year={2013},
  publisher={The American Astronomical Society},
  doi={10.1088/0004-637X/774/1/51}
}

@article{mutschler2015freeze,
  title={Freeze--thaw cycles as drivers of complex ribozyme assembly},
  author={Mutschler, Hannes and Wochner, Aniela and Holliger, Philipp},
  journal={Nature Chemistry},
  volume={7},
  number={6},
  pages={502--508},
  year={2015},
  publisher={Nature Publishing Group UK London},
  doi={10.1038/nchem.2251}
}

@article{attwater2025trinucleotide,
  title={Trinucleotide substrates under pH--freeze--thaw cycles enable open-ended exponential RNA replication by a polymerase ribozyme},
  author={Attwater, James and Augustin, Teresa L and Curran, Joseph F and Kwok, Samantha LY and Ohlendorf, Luis and Gianni, Edoardo and Holliger, Philipp},
  journal={Nature Chemistry},
  volume={17},
  number={7},
  pages={1129--1137},
  year={2025},
  publisher={Nature Publishing Group UK London},
  doi={10.1038/s41557-025-01830-y}
}

@article{zhang2022freeze,
  author = {Zhang, Stephanie J. and Duzdevich, Daniel and Ding, Dian and Szostak, Jack W.},
  title = {Freeze-thaw cycles enable a prebiotically plausible and continuous pathway from nucleotide activation to nonenzymatic RNA copying},
  journal = {Proceedings of the National Academy of Sciences},
  volume = {119},
  number = {17},
  pages = {e2116429119},
  year = {2022},
  doi = {10.1073/pnas.2116429119}
}

@article{shields2013effect,
  title={The effect of host star spectral energy distribution and ice-albedo feedback on the climate of extrasolar planets},
  author={Shields, Aomawa L and Meadows, Victoria S and Bitz, Cecilia M and Pierrehumbert, Raymond T and Joshi, Manoj M and Robinson, Tyler D},
  journal={Astrobiology},
  volume={13},
  number={8},
  pages={715--739},
  year={2013},
  publisher={SAGE Publications Sage CA: Los Angeles, CA},
  doi={10.1089/ast.2012.0961}
}

@article{joshi2012suppression,
  title={Suppression of the water ice and snow albedo feedback on planets orbiting red dwarf stars and the subsequent widening of the habitable zone},
  author={Joshi, Manoj M and Haberle, Robert M},
  journal={Astrobiology},
  volume={12},
  number={1},
  pages={3--8},
  year={2012},
  publisher={SAGE Publications Sage CA: Los Angeles, CA},
  doi={10.1089/ast.2011.0668}
}

@article{shields2016habitability,
  title={The habitability of planets orbiting M-dwarf stars},
  author={Shields, Aomawa L and Ballard, Sarah and Johnson, John Asher},
  journal={Physics Reports},
  volume={663},
  pages={1--38},
  year={2016},
  publisher={Elsevier},
  doi={10.1016/j.physrep.2016.10.003}
}

@article{labonte2020sensitivity,
  title={Sensitivity of the atmospheric water cycle within the habitable zone of a tidally locked, Earth-like exoplanet},
  author={Labont{\'e}, Marie-Pier and Merlis, Timothy M},
  journal={The Astrophysical Journal},
  volume={896},
  number={1},
  pages={31},
  year={2020},
  publisher={The American Astronomical Society},
  doi={10.3847/1538-4357/ab9102}
}

@article{hourdin2006lmdz4,
  title={The LMDZ4 general circulation model: climate performance and sensitivity to parametrized physics with emphasis on tropical convection},
  author={Hourdin, Fr{\'e}d{\'e}ric and Musat, Ionela and Bony, Sandrine and Braconnot, Pascale and Codron, Francis and Dufresne, Jean-Louis and Fairhead, Laurent and Filiberti, Marie-Ang{\`e}le and Friedlingstein, Pierre and Grandpeix, Jean-Yves and others},
  journal={Climate Dynamics},
  volume={27},
  number={7},
  pages={787--813},
  year={2006},
  publisher={Springer},
  doi={10.1007/s00382-006-0158-0}
}

@article{anglada2016terrestrial,
  title={A terrestrial planet candidate in a temperate orbit around Proxima Centauri},
  author={Anglada-Escud{\'e}, Guillem and Amado, Pedro J and Barnes, John and Berdi{\~n}as, Zaira M and Butler, R Paul and Coleman, Gavin AL and de La Cueva, Ignacio and Dreizler, Stefan and Endl, Michael and Giesers, Benjamin and others},
  journal={nature},
  volume={536},
  number={7617},
  pages={437--440},
  year={2016},
  publisher={Nature Publishing Group UK London},
  doi={10.1038/nature19106}
}

@article{ribas2016habitability,
  title={The habitability of Proxima Centauri b: I. Irradiation, rotation and volatile inventory from formation to the present},
  author={Ribas, Ignasi and Bolmont, Emeline and Selsis, Franck and Reiners, Ansgar and Leconte, J{\'e}r{\'e}my and Raymond, Sean N and Engle, Scott G and Guinan, Edward F and Morin, Julien and Turbet, Martin and others},
  journal={Astronomy \& Astrophysics},
  volume={596},
  pages={A111},
  year={2016},
  publisher={EDP Sciences},
  doi={10.1051/0004-6361/201629576}
}

@article{gillon2016temperate,
  title={Temperate Earth-sized planets transiting a nearby ultracool dwarf star},
  author={Gillon, Micha{\"e}l and Jehin, Emmanu{\"e}l and Lederer, Susan M and Delrez, Laetitia and de Wit, Julien and Burdanov, Artem and Van Grootel, Val{\'e}rie and Burgasser, Adam J and Triaud, Amaury HMJ and Opitom, Cyrielle and others},
  journal={Nature},
  volume={533},
  number={7602},
  pages={221--224},
  year={2016},
  publisher={Nature Publishing Group UK London},
  doi={10.1038/nature17448}
}

@article{gillon2017seven,
  title={Seven temperate terrestrial planets around the nearby ultracool dwarf star TRAPPIST-1},
  author={Gillon, Micha{\"e}l and Triaud, Amaury HMJ and Demory, Brice-Olivier and Jehin, Emmanu{\"e}l and Agol, Eric and Deck, Katherine M and Lederer, Susan M and De Wit, Julien and Burdanov, Artem and Ingalls, James G and others},
  journal={Nature},
  volume={542},
  number={7642},
  pages={456--460},
  year={2017},
  publisher={Nature Publishing Group UK London},
  doi={10.1038/nature21360}
}

@article{kasting1993habitable,
  title={Habitable zones around main sequence stars},
  author={Kasting, James F and Whitmire, Daniel P and Reynolds, Ray T},
  journal={Icarus},
  volume={101},
  number={1},
  pages={108--128},
  year={1993},
  publisher={Elsevier},
  doi={10.1006/icar.1993.1010}
}

@article{des2008nasa,
  title={The NASA astrobiology roadmap},
  author={Des Marais, David J and Nuth III, Joseph A and Allamandola, Louis J and Boss, Alan P and Farmer, Jack D and Hoehler, Tori M and Jakosky, Bruce M and Meadows, Victoria S and Pohorille, Andrew and Runnegar, Bruce and others},
  journal={Astrobiology},
  volume={8},
  number={4},
  pages={715--730},
  year={2008},
  publisher={SAGE Publications Sage CA: Los Angeles, CA},
  doi={10.1089/ast.2008.0819}
}

@article{cockell2016habitability,
  title={Habitability: a review},
  author={Cockell, Charles S and Bush, Tim and Bryce, Casey and Direito, Susana and Fox-Powell, Mark and Harrison, Jesse Patrick and Lammer, Helmut and Landenmark, H and Martin-Torres, Javier and Nicholson, N and others},
  journal={Astrobiology},
  volume={16},
  number={1},
  pages={89--117},
  year={2016},
  publisher={SAGE Publications Sage CA: Los Angeles, CA},
  doi={10.1089/ast.2015.1295}
}

@article{brack1993liquid,
  title={Liquid water and the origin of life},
  author={Brack, Andr{\'e}},
  journal={Origins of Life and Evolution of the Biosphere},
  volume={23},
  number={1},
  pages={3--10},
  year={1993},
  publisher={Springer},
  doi={10.1007/BF01581985}
}

@article{alibert2014radius,
  title={On the radius of habitable planets},
  author={Alibert, Yann},
  journal={Astronomy \& Astrophysics},
  volume={561},
  pages={A41},
  year={2014},
  publisher={EDP Sciences},
  doi={10.1051/0004-6361/201322293}
}

@article{dittmann2017temperate,
  title={A temperate rocky super-Earth transiting a nearby cool star},
  author={Dittmann, Jason A and Irwin, Jonathan M and Charbonneau, David and Bonfils, Xavier and Astudillo-Defru, Nicola and Haywood, Rapha{\"e}lle D and Berta-Thompson, Zachory K and Newton, Elisabeth R and Rodriguez, Joseph E and Winters, Jennifer G and others},
  journal={Nature},
  volume={544},
  number={7650},
  pages={333--336},
  year={2017},
  publisher={Nature Publishing Group UK London},
  doi={10.1038/nature22055}
}

@article{barnes2017tidal,
  title={Tidal locking of habitable exoplanets},
  author={Barnes, Rory},
  journal={Celestial mechanics and dynamical astronomy},
  volume={129},
  number={4},
  pages={509--536},
  year={2017},
  publisher={Springer},
  doi={10.1007/s10569-017-9783-7}
}

@article{turbet2018modeling,
  title={Modeling climate diversity, tidal dynamics and the fate of volatiles on TRAPPIST-1 planets},
  author={Turbet, Martin and Bolmont, Emeline and Leconte, Jeremy and Forget, Fran{\c{c}}ois and Selsis, Franck and Tobie, Gabriel and Caldas, Anthony and Naar, Joseph and Gillon, Micha{\"e}l},
  journal={Astronomy \& Astrophysics},
  volume={612},
  pages={A86},
  year={2018},
  publisher={EDP Sciences},
  doi={10.1051/0004-6361/201731620}
}

@article{joshi1997simulations,
  title={Simulations of the atmospheres of synchronously rotating terrestrial planets orbiting M dwarfs: conditions for atmospheric collapse and the implications for habitability},
  author={Joshi, MM and Haberle, RM and Reynolds, RT},
  journal={Icarus},
  volume={129},
  number={2},
  pages={450--465},
  year={1997},
  publisher={Elsevier},
  doi={10.1006/icar.1997.5793}
}

@article{joshi2003climate,
  title={Climate model studies of synchronously rotating planets},
  author={Joshi, Manoj},
  journal={Astrobiology},
  volume={3},
  number={2},
  pages={415--427},
  year={2003},
  publisher={SAGE Publications Sage CA: Los Angeles, CA},
  doi={doi.org/10.1089/153110703769016488}
}

@article{haqq2018demarcating,
  title={Demarcating circulation regimes of synchronously rotating terrestrial planets within the habitable zone},
  author={Haqq-Misra, Jacob and Wolf, Eric T and Joshi, Manoj and Zhang, Xi and Kopparapu, Ravi Kumar},
  journal={The Astrophysical Journal},
  volume={852},
  number={2},
  pages={67},
  year={2018},
  publisher={The American Astronomical Society},
  doi={10.3847/1538-4357/aa9f1f}
}

@article{lefevre20213d,
  title={3D convection-resolving model of temperate, tidally locked exoplanets},
  author={Lef{\`e}vre, Maxence and Turbet, Martin and Pierrehumbert, Raymond},
  journal={The Astrophysical Journal},
  volume={913},
  number={2},
  pages={101},
  year={2021},
  publisher={The American Astronomical Society},
  doi={10.3847/1538-4357/abf2c1}
}

@article{sergeev2022bistability,
  title={Bistability of the atmospheric circulation on TRAPPIST-1e},
  author={Sergeev, Denis E and Lewis, Neil T and Lambert, F Hugo and Mayne, Nathan J and Boutle, Ian A and Manners, James and Kohary, Krisztian},
  journal={The Planetary Science Journal},
  volume={3},
  number={9},
  pages={214},
  year={2022},
  publisher={The American Astronomical Society},
  doi={10.3847/PSJ/ac83be}
}

@article{taniguchi2026atmospheric,
  title={Atmospheric collapse and habitability on tidally locked exoplanets},
  author={Taniguchi, Keigo and Kodama, Takanori and Turbet, Martin and Chaverot, Guillaume and Millour, Ehouarn and Genda, Hidenori},
  journal={The Astrophysical Journal},
  volume={1000},
  number={1},
  pages={99},
  year={2026},
  publisher={The American Astronomical Society},
  doi={10.3847/1538-4357/ae47f2}
}

@article{wordsworth2011gliese,
  title={Gliese 581d is the first discovered terrestrial-mass exoplanet in the habitable zone},
  author={Wordsworth, Robin D and Forget, Fran{\c{c}}ois and Selsis, Franck and Millour, Ehouarn and Charnay, Benjamin and Madeleine, Jean-Baptiste},
  journal={The Astrophysical Journal Letters},
  volume={733},
  number={2},
  pages={L48},
  year={2011},
  publisher={The American Astronomical Society},
  doi={10.1088/2041-8205/733/2/L48}
}

@article{yang2013stabilizing,
  title={Stabilizing cloud feedback dramatically expands the habitable zone of tidally locked planets},
  author={Yang, Jun and Cowan, Nicolas B and Abbot, Dorian S},
  journal={The Astrophysical Journal Letters},
  volume={771},
  number={2},
  pages={L45},
  year={2013},
  publisher={The American Astronomical Society},
  doi={10.1088/2041-8205/771/2/L45}
}

@article{hu2014role,
  title={Role of ocean heat transport in climates of tidally locked exoplanets around M dwarf stars},
  author={Hu, Yongyun and Yang, Jun},
  journal={Proceedings of the National Academy of Sciences},
  volume={111},
  number={2},
  pages={629--634},
  year={2014},
  publisher={National Academy of Sciences},
  doi={10.1073/pnas.1315215111}
}

@article{heath1999habitability,
  title={Habitability of planets around red dwarf stars},
  author={Heath, Martin J and Doyle, Laurance R and Joshi, Manoj M and Haberle, Robert M},
  journal={Origins of Life and Evolution of the Biosphere},
  volume={29},
  number={4},
  pages={405--424},
  year={1999},
  publisher={Springer},
  doi={10.1023/A:1006596718708}
}

@article{wolf2019simulated,
  title={Simulated phase-dependent spectra of terrestrial Aquaplanets in m dwarf systems},
  author={Wolf, Eric T and Kopparapu, Ravi K and Haqq-Misra, Jacob},
  journal={The Astrophysical Journal},
  volume={877},
  number={1},
  pages={35},
  year={2019},
  publisher={The American Astronomical Society},
  doi={10.3847/1538-4357/ab184a}
}

@article{tarter2007reappraisal,
  title={A reappraisal of the habitability of planets around M dwarf stars},
  author={Tarter, Jill C and Backus, Peter R and Mancinelli, Rocco L and Aurnou, Jonathan M and Backman, Dana E and Basri, Gibor S and Boss, Alan P and Clarke, Andrew and Deming, Drake and Doyle, Laurance R and others},
  journal={Astrobiology},
  volume={7},
  number={1},
  pages={30--65},
  year={2007},
  publisher={SAGE Publications Sage CA: Los Angeles, CA},
  doi={10.1089/ast.2006.0124}
}

@article{leconte20133d,
  title={3D climate modeling of close-in land planets: circulation patterns, climate moist bistability, and habitability},
  author={Leconte, J{\'e}r{\'e}my and Forget, Francois and Charnay, Benjamin and Wordsworth, Robin and Selsis, Franck and Millour, Ehouarn and Spiga, Aymeric},
  journal={Astronomy \& Astrophysics},
  volume={554},
  pages={A69},
  year={2013},
  publisher={EDP Sciences},
  doi={10.1051/0004-6361/201321042}
}

@article{del2019habitable,
  title={Habitable climate scenarios for Proxima Centauri b with a dynamic ocean},
  author={Del Genio, Anthony D and Way, Michael J and Amundsen, David S and Aleinov, Igor and Kelley, Maxwell and Kiang, Nancy Y and Clune, Thomas L},
  journal={Astrobiology},
  volume={19},
  number={1},
  pages={99--125},
  year={2019},
  doi={10.1089/ast.2017.1760}
}

@article{peacock2019predicting,
  title={Predicting the extreme ultraviolet radiation environment of exoplanets around low-mass stars: the TRAPPIST-1 system},
  author={Peacock, Sarah and Barman, Travis and Shkolnik, Evgenya L and Hauschildt, Peter H and Baron, E},
  journal={The Astrophysical Journal},
  volume={871},
  number={2},
  pages={235},
  year={2019},
  publisher={The American Astronomical Society},
  doi={10.3847/1538-4357/aaf891}
}

@article{abbot2012indication,
  title={Indication of insensitivity of planetary weathering behavior and habitable zone to surface land fraction},
  author={Abbot, Dorian S and Cowan, Nicolas B and Ciesla, Fred J},
  journal={The Astrophysical Journal},
  volume={756},
  number={2},
  pages={178},
  year={2012},
  publisher={The American Astronomical Society},
  doi={10.1088/0004-637X/756/2/178}
}

@article{foley2015role,
  title={The role of plate tectonic--climate coupling and exposed land area in the development of habitable climates on rocky planets},
  author={Foley, Bradford J},
  journal={The Astrophysical Journal},
  volume={812},
  number={1},
  pages={36},
  year={2015},
  publisher={The American Astronomical Society},
  doi={10.1088/0004-637X/812/1/36}
}

@article{foley2019habitability,
  title={Habitability of Earth-like stagnant lid planets: climate evolution and recovery from snowball states},
  author={Foley, Bradford J},
  journal={The Astrophysical Journal},
  volume={875},
  number={1},
  pages={72},
  year={2019},
  publisher={The American Astronomical Society},
  doi={10.3847/1538-4357/ab0f31}
}

@article{yang2014water,
  title={Water trapping on tidally locked terrestrial planets requires special conditions},
  author={Yang, Jun and Liu, Yonggang and Hu, Yongyun and Abbot, Dorian S},
  journal={The Astrophysical Journal Letters},
  volume={796},
  number={2},
  pages={L22},
  year={2014},
  publisher={The American Astronomical Society},
  doi={10.1088/2041-8205/796/2/L22}
}

@article{turbet2016habitability,
  title={The habitability of Proxima Centauri b-II. Possible climates and observability},
  author={Turbet, Martin and Leconte, J{\'e}r{\'e}my and Selsis, Franck and Bolmont, Emeline and Forget, Fran{\c{c}}ois and Ribas, Ignasi and Raymond, Sean N and Anglada-Escud{\'e}, Guillem},
  journal={Astronomy \& Astrophysics},
  volume={596},
  pages={A112},
  year={2016},
  publisher={EDP Sciences},
  doi={10.1051/0004-6361/201629577}
}

@article{ojha2022liquid,
  title={Liquid water on cold exo-Earths via basal melting of ice sheets},
  author={Ojha, Lujendra and Troncone, Bryce and Buffo, Jacob and Journaux, Baptiste and McDonald, George},
  journal={Nature communications},
  volume={13},
  number={1},
  pages={7521},
  year={2022},
  publisher={Nature Publishing Group UK London},
  doi={10.1038/s41467-022-35187-4}
}

@article{makarov2012dynamical,
  title={Dynamical evolution and spin--orbit resonances of potentially habitable exoplanets: the case of GJ 581d},
  author={Makarov, Valeri V and Berghea, Ciprian and Efroimsky, Michael},
  journal={The Astrophysical Journal},
  volume={761},
  number={2},
  pages={83},
  year={2012},
  publisher={The American Astronomical Society},
  doi={10.1088/0004-637X/761/2/83}
}

@article{rodriguez2012spin,
  title={Spin--orbit coupling for tidally evolving super-Earths},
  author={Rodr{\'\i}guez, Adri{\'a}n and Callegari Jr, Nelson and Michtchenko, Tatiana Alexandrovna and Hussmann, Hauke},
  journal={Monthly Notices of the Royal Astronomical Society},
  volume={427},
  number={3},
  pages={2239--2250},
  year={2012},
  publisher={Blackwell Science Ltd Oxford, UK},
  doi={10.1111/j.1365-2966.2012.22084.x}
}

@article{wang2014climate,
  title={Climate patterns of habitable exoplanets in eccentric orbits around M dwarfs},
  author={Wang, Yuwei and Tian, Feng and Hu, Yongyun},
  journal={The Astrophysical Journal Letters},
  volume={791},
  number={1},
  pages={L12},
  year={2014},
  publisher={The American Astronomical Society},
  doi={10.1088/2041-8205/791/1/L12}
}

@article{shakespeare2023day,
  title={Day and night: habitability of tidally locked planets with sporadic rotation},
  author={Shakespeare, Cody J and Steffen, Jason H},
  journal={Monthly Notices of the Royal Astronomical Society},
  volume={524},
  number={4},
  pages={5708--5724},
  year={2023},
  publisher={Oxford University Press},
  doi={10.1093/mnras/stad2162}
}

@article{chen2023sporadic,
  title={Sporadic spin-orbit variations in compact multiplanet systems and their influence on exoplanet climate},
  author={Chen, Howard and Li, Gongjie and Paradise, Adiv and Kopparapu, Ravi K},
  journal={The Astrophysical Journal Letters},
  volume={946},
  number={2},
  pages={L32},
  year={2023},
  publisher={The American Astronomical Society},
  doi={10.3847/2041-8213/acbd33}
}

@article{revol2024drifts,
  title={Drifts of the substellar points of the TRAPPIST-1 planets},
  author={Revol, Alexandre and Bolmont, {\'E}meline and Sastre, Mariana and Tobie, Gabriel and Libert, Anne-Sophie and Kervazo, Mathilde and Blanco-Cuaresma, Sergi},
  journal={Astronomy \& Astrophysics},
  volume={691},
  pages={L3},
  year={2024},
  publisher={EDP Sciences},
  doi={10.1051/0004-6361/202451532}
}

@article{vinson2019chaotic,
  title={The chaotic nature of TRAPPIST-1 planetary spin states},
  author={Vinson, Alec M and Tamayo, Daniel and Hansen, Brad MS},
  journal={Monthly Notices of the Royal Astronomical Society},
  volume={488},
  number={4},
  pages={5739--5747},
  year={2019},
  publisher={Oxford University Press},
  doi={10.1093/mnras/stz2113}
}

@article{shields2014spectrum,
  title={Spectrum-driven planetary deglaciation due to increases in stellar luminosity},
  author={Shields, Aomawa L and Bitz, Cecilia M and Meadows, Victoria S and Joshi, Manoj M and Robinson, Tyler D},
  journal={The Astrophysical Journal Letters},
  volume={785},
  number={1},
  pages={L9},
  year={2014},
  publisher={The American Astronomical Society},
  doi={10.1088/2041-8205/785/1/L9}
}

@article{forget20133d,
  title={3D modelling of the early martian climate under a denser CO2 atmosphere: Temperatures and CO2 ice clouds},
  author={Forget, Fran{\c{c}}ois and Wordsworth, Robin and Millour, Ehouarn and Madeleine, J-B and Kerber, Laura and Leconte, J{\'e}r{\'e}my and Marcq, Emmanuel and Haberle, Robert M},
  journal={Icarus},
  volume={222},
  number={1},
  pages={81--99},
  year={2013},
  publisher={Elsevier},
  doi={10.1016/j.icarus.2012.10.019}
}

@article{charnay2014titan,
  title={Titan’s past and future: 3D modeling of a pure nitrogen atmosphere and geological implications},
  author={Charnay, Benjamin and Forget, Fran{\c{c}}ois and Tobie, Gabriel and Sotin, Christophe and Wordsworth, Robin},
  journal={Icarus},
  volume={241},
  pages={269--279},
  year={2014},
  publisher={Elsevier},
  doi={10.1016/j.icarus.2014.07.009}
}

@article{turbet2021day,
  title={Day--night cloud asymmetry prevents early oceans on Venus but not on Earth},
  author={Turbet, Martin and Bolmont, Emeline and Chaverot, Guillaume and Ehrenreich, David and Leconte, J{\'e}r{\'e}my and Marcq, Emmanuel},
  journal={Nature},
  volume={598},
  number={7880},
  pages={276--280},
  year={2021},
  publisher={Nature Publishing Group UK London},
  doi={10.1038/s41586-021-03873-w}
}

@article{walker1981negative,
  title={A negative feedback mechanism for the long-term stabilization of Earth's surface temperature},
  author={Walker, James CG and Hays, Phillip B and Kasting, James F},
  journal={Journal of Geophysical Research: Oceans},
  volume={86},
  number={C10},
  pages={9776--9782},
  year={1981},
  publisher={Wiley Online Library},
  doi={10.1029/JC086iC10p09776}
}

@article{bolmont2016habitability,
  title={Habitability of planets on eccentric orbits: Limits of the mean flux approximation},
  author={Bolmont, Emeline and Libert, Anne-Sophie and Leconte, Jeremy and Selsis, Franck},
  journal={Astronomy \& Astrophysics},
  volume={591},
  pages={A106},
  year={2016},
  publisher={EDP Sciences},
  doi={10.1051/0004-6361/201628073}
}

@article{sergeev2022trappist,
  title={The TRAPPIST-1 habitable atmosphere intercomparison (THAI). II. Moist cases—The two waterworlds},
  author={Sergeev, Denis E and Fauchez, Thomas J and Turbet, Martin and Boutle, Ian A and Tsigaridis, Kostas and Way, Michael J and Wolf, Eric T and Domagal-Goldman, Shawn D and Forget, Fran{\c{c}}ois and Haqq-Misra, Jacob and others},
  journal={The Planetary Science Journal},
  volume={3},
  number={9},
  pages={212},
  year={2022},
  publisher={The American Astronomical Society},
  doi={10.3847/PSJ/ac6cf2}
}

@article{leconte2013increased,
  title={Increased insolation threshold for runaway greenhouse processes on Earth-like planets},
  author={Leconte, J{\'e}r{\'e}my and Forget, Francois and Charnay, Benjamin and Wordsworth, Robin and Pottier, Aliz{\'e}e},
  journal={Nature},
  volume={504},
  number={7479},
  pages={268--271},
  year={2013},
  publisher={Nature Publishing Group UK London},
  doi={10.1038/nature12827}
}

@article{agol2021refining,
  title={Refining the transit-timing and photometric analysis of TRAPPIST-1: masses, radii, densities, dynamics, and ephemerides},
  author={Agol, Eric and Dorn, Caroline and Grimm, Simon L and Turbet, Martin and Ducrot, Elsa and Delrez, Laetitia and Gillon, Micha{\"e}l and Demory, Brice-Olivier and Burdanov, Artem and Barkaoui, Khalid and others},
  journal={The planetary science journal},
  volume={2},
  number={1},
  pages={1},
  year={2021},
  publisher={The American Astronomical Society},
  doi={10.3847/PSJ/abd0221}
}

@article{turbet2022trappist,
  title={The TRAPPIST-1 Habitable Atmosphere Intercomparison (THAI). I. Dry Cases—The Fellowship of the GCMs},
  author={Turbet, Martin and Fauchez, Thomas J and Sergeev, Denis E and Boutle, Ian A and Tsigaridis, Kostas and Way, Michael J and Wolf, Eric T and Domagal-Goldman, Shawn D and Forget, Fran{\c{c}}ois and Haqq-Misra, Jacob and others},
  journal={The Planetary Science Journal},
  volume={3},
  number={9},
  pages={211},
  year={2022},
  publisher={The American Astronomical Society},
  doi={10.3847/PSJ/ac6cf0}
}

@article{charette2010volume,
  title={The volume of Earth's ocean},
  author={Charette, Matthew A and Smith, Walter HF},
  journal={Oceanography},
  volume={23},
  number={2},
  pages={112--114},
  year={2010},
  publisher={JSTOR},
  doi={10.5670/oceanog.2010.51}
}

@article{barletta2018observed,
  title={Observed rapid bedrock uplift in Amundsen Sea Embayment promotes ice-sheet stability},
  author={Barletta, Valentina R and Bevis, Michael and Smith, Benjamin E and Wilson, Terry and Brown, Abel and Bordoni, Andrea and Willis, Michael and Khan, Shfaqat Abbas and Rovira-Navarro, Marc and Dalziel, Ian and others},
  journal={Science},
  volume={360},
  number={6395},
  pages={1335--1339},
  year={2018},
  publisher={American Association for the Advancement of Science},
  doi={10.1126/science.aao1447}
}

@article{westall2018importance,
  title={The importance of water for life},
  author={Westall, Frances and Brack, Andr{\'e}},
  journal={Space Science Reviews},
  volume={214},
  number={2},
  pages={50},
  year={2018},
  publisher={Springer},
  doi={10.1007/s11214-018-0476-7}
}

@misc{petrenko2002physics,
  title={Physics of Ice. Oxford Univ},
  author={Petrenko, V and Whitworth, RW},
  year={2002},
  publisher={Press},
  doi={10.1093/acprof:oso/9780198518945.001.0001}
}

@article{wolf2017assessing,
  title={Assessing the habitability of the TRAPPIST-1 system using a 3D climate model},
  author={Wolf, Eric T},
  journal={The Astrophysical Journal Letters},
  volume={839},
  number={1},
  pages={L1},
  year={2017},
  publisher={The American Astronomical Society},
  doi={10.3847/2041-8213/aa693a}
}

@article{abe1993physical,
  title={Physical state of the very early Earth},
  author={Abe, Yutaka},
  journal={Lithos},
  volume={30},
  number={3-4},
  pages={223--235},
  year={1993},
  publisher={Elsevier},
  doi={10.1016/0024-4937(93)90037-D}
}

@article{abbot2011steppenwolf,
  title={The Steppenwolf: a proposal for a habitable planet in interstellar space},
  author={Abbot, Dorian S and Switzer, Eric R},
  journal={The Astrophysical Journal Letters},
  volume={735},
  number={2},
  pages={L27},
  year={2011},
  publisher={The American Astronomical Society},
  doi={10.1088/2041-8205/735/2/L27}
}

@article{whitehouse2018glacial,
  title={Glacial isostatic adjustment modelling: historical perspectives, recent advances, and future directions},
  author={Whitehouse, Pippa L},
  journal={Earth surface dynamics},
  volume={6},
  number={2},
  pages={401--429},
  year={2018},
  publisher={Copernicus Publications G{\"o}ttingen, Germany},
  doi={10.5194/esurf-6-401-2018}
}

@article{lincowski2018evolved,
  title={Evolved climates and observational discriminants for the TRAPPIST-1 planetary system},
  author={Lincowski, Andrew P and Meadows, Victoria S and Crisp, David and Robinson, Tyler D and Luger, Rodrigo and Lustig-Yaeger, Jacob and Arney, Giada N},
  journal={The Astrophysical Journal},
  volume={867},
  number={1},
  pages={76},
  year={2018},
  publisher={The American Astronomical Society},
  doi={10.3847/1538-4357/aae36a}
}

@article{davies2010earth,
  title={Earth's surface heat flux},
  author={Davies, J Huw and Davies, D Rhodri},
  journal={Solid earth},
  volume={1},
  number={1},
  pages={5--24},
  year={2010},
  publisher={Copernicus Publications G{\"o}ttingen, Germany},
  doi={10.5194/se-1-5-2010}
}

@article{airy1855,
  title={On the Computation of the Effect of the Attraction of Mountain-Masses, as Disturbing the Apparent Astronomical Latitude of Stations in Geodetic Surveys},
  author={Airy, George Biddell},
  journal={Philosophical Transactions of the Royal Society of London},
  volume={145},
  pages={101--104},
  year={1855},
  doi={10.1098/rstl.1855.0003}
}

@article{dobos2019tidal,
  title={Tidal heating and the habitability of the TRAPPIST-1 exoplanets},
  author={Dobos, Vera and Barr, Amy C and Kiss, L{\'a}szl{\'o} L},
  journal={Astronomy \& Astrophysics},
  volume={624},
  pages={A2},
  year={2019},
  publisher={EDP Sciences},
  doi={10.1051/0004-6361/201834254}
}

@article{barr2018interior,
  title={Interior structures and tidal heating in the TRAPPIST-1 planets},
  author={Barr, Amy C and Dobos, Vera and Kiss, L{\'a}szl{\'o} L},
  journal={Astronomy \& Astrophysics},
  volume={613},
  pages={A37},
  year={2018},
  publisher={EDP Sciences},
  doi={10.1051/0004-6361/201731992}
}

@article{bolmont2026estimation,
  title={Estimation of the tidal heating in the TRAPPIST-1 planets: Influence of the internal structure},
  author={Bolmont, Emeline and Sastre, Mariana and Revol, Alexandre and Kervazo, Mathilde and Tobie, Gabriel},
  journal={Astronomy \& Astrophysics},
  volume={707},
  pages={A101},
  year={2026},
  publisher={EDP Sciences},
  doi={10.1051/0004-6361/202554529}
}

@article{paradise2022exoplasim,
  title={ExoPlaSim: Extending the planet simulator for exoplanets},
  author={Paradise, Adiv and Macdonald, Evelyn and Menou, Kristen and Lee, Christopher and Fan, Bo Lin},
  journal={Monthly Notices of the Royal Astronomical Society},
  volume={511},
  number={3},
  pages={3272--3303},
  year={2022},
  publisher={Oxford University Press},
  doi={10.1093/mnras/stac172}
}

@article{bhatnagar2026fast,
  title={A fast and physically grounded ocean model for GCMs: the Dynamical Slab Ocean Model of the Generic-PCM (rev. 3423)},
  author={Bhatnagar, S. and Codron, F. and Millour, E. and Bolmont, E. and Brunetti, M. and Kasparian, J. and Turbet, M. and Chaverot, G.},
  journal={Geoscientific Model Development},
  volume={19},
  pages={3285--3316},
  year={2026},
  publisher={Copernicus Publications G{\"o}ttingen, Germany},
  doi={10.5194/gmd-19-3285-2026}
}

@article{fu1992correlated,
  title={On the correlated k-distribution method for radiative transfer in nonhomogeneous atmospheres},
  author={Fu, Qiang and Liou, KN},
  journal={Journal of Atmospheric Sciences},
  volume={49},
  number={22},
  pages={2139--2156},
  year={1992},
  doi={10.1175/1520-0469(1992)049<2139:OTCDMF>2.0.CO;2}
}

@article{eymet2016kspectrum,
  title={kspectrum: An open-source code for high-resolution molecular absorption spectra production},
  author={Eymet, Vincent and Coustet, C{\'e}dric and Piaud, Beno{\^\i}t},
  journal={Journal of Physics: Conference Series},
  volume={676},
  number={1},
  pages={012005},
  year={2016},
  organization={IOP Publishing},
  doi={10.1088/1742-6596/676/1/012005}
}

@article{charnay2013exploring,
  title={Exploring the faint young Sun problem and the possible climates of the Archean Earth with a 3-D GCM},
  author={Charnay, Benjamin and Forget, Fran{\c{c}}ois and Wordsworth, Robin and Leconte, J{\'e}r{\'e}my and Millour, Ehouarn and Codron, Francis and Spiga, Aymeric},
  journal={Journal of Geophysical Research: Atmospheres},
  volume={118},
  number={18},
  pages={10--414},
  year={2013},
  publisher={Wiley Online Library},
  doi={10.1002/jgrd.50808}
}

@article{merlis2010atmospheric,
  title={Atmospheric dynamics of Earth-like tidally locked Aquaplanets},
  author={Merlis, Timothy M and Schneider, Tapio},
  journal={Journal of Advances in Modeling Earth Systems},
  volume={2},
  number={4},
  year={2010},
  publisher={Wiley Online Library},
  doi={10.3894/JAMES.2010.2.13}
}

@article{haqq2022energy,
  title={An energy balance model for rapidly and synchronously rotating terrestrial planets},
  author={Haqq-Misra, Jacob and Hayworth, Benjamin PC},
  journal={The Planetary Science Journal},
  volume={3},
  number={2},
  pages={32},
  year={2022},
  publisher={The American Astronomical Society},
  doi={10.3847/PSJ/ac49eb}
}

@article{henry2006solar,
  title={The solar neighborhood. XVII. Parallax results from the CTIOPI 0.9 m program: 20 new members of the RECONS 10 parsec sample},
  author={Henry, Todd J and Jao, Wei-Chun and Subasavage, John P and Beaulieu, Thomas D and Ianna, Philip A and Costa, Edgardo and M{\'e}ndez, Ren{\'e} A},
  journal={The Astronomical Journal},
  volume={132},
  number={6},
  pages={2360--2371},
  year={2006},
  doi={10.1086/508233}
}

@article{maurel2025constraints,
  title={Constraints on the possible atmospheres on TRAPPIST-1 b: insights from 3D climate modeling},
  author={Maurel, Alice and Turbet, Martin and Ducrot, Elsa and Leconte, J{\'e}r{\'e}my and Chaverot, Guillaume and Milcareck, Gwenael and Revol, Alexandre and Charnay, Benjamin and Fauchez, Thomas J and Gillon, Micha{\"e}l and others},
  journal={Astronomy \& Astrophysics},
  volume={701},
  pages={A193},
  year={2025},
  publisher={EDP Sciences},
  doi={10.1051/0004-6361/202554243}
}

@article{heller2011tidal,
  title={Tidal obliquity evolution of potentially habitable planets},
  author={Heller, R{\'e}ne and Leconte, J{\'e}r{\'e}my and Barnes, Rory},
  journal={Astronomy \& Astrophysics},
  volume={528},
  pages={A27},
  year={2011},
  publisher={EDP Sciences},
  doi={10.1051/0004-6361/201015809}
}

@article{henry2024character,
  title={The character of m dwarfs},
  author={Henry, Todd J and Jao, Wei-Chun},
  journal={Annual Review of Astronomy and Astrophysics},
  volume={62},
  number={1},
  pages={593--633},
  year={2024},
  publisher={Annual Reviews},
  doi={10.1146/annurev-astro-052722-102740}
}

@article{kopparapu2013habitable,
  title={Habitable zones around main-sequence stars: new estimates},
  author={Kopparapu, Ravi Kumar and Ramirez, Ramses and Kasting, James F and Eymet, Vincent and Robinson, Tyler D and Mahadevan, Suvrath and Terrien, Ryan C and Domagal-Goldman, Shawn and Meadows, Victoria and Deshpande, Rohit},
  journal={The Astrophysical Journal},
  volume={765},
  number={2},
  pages={131},
  year={2013},
  publisher={The American Astronomical Society},
  doi={10.1088/0004-637X/765/2/131}
}

@article{kopparapu2014habitable,
  title={Habitable zones around main-sequence stars: dependence on planetary mass},
  author={Kopparapu, Ravi Kumar and Ramirez, Ramses M and SchottelKotte, James and Kasting, James F and Domagal-Goldman, Shawn and Eymet, Vincent},
  journal={The Astrophysical Journal Letters},
  volume={787},
  number={2},
  pages={L29},
  year={2014},
  publisher={The American Astronomical Society},
  doi={10.1088/2041-8205/787/2/L29}
}

@article{kopparapu2017habitable,
  title={Habitable moist atmospheres on terrestrial planets near the inner edge of the habitable zone around M dwarfs},
  author={Kopparapu, Ravi Kumar and Wolf, Eric T and Arney, Giada and Batalha, Natasha E and Haqq-Misra, Jacob and Grimm, Simon L and Heng, Kevin},
  journal={The Astrophysical Journal},
  volume={845},
  number={1},
  pages={5},
  year={2017},
  publisher={The American Astronomical Society},
  doi={10.3847/1538-4357/aa7cf9}
}

@article{bardet2022saturn,
       author = {{Bardet}, Deborah and {Spiga}, Aymeric and {Guerlet}, Sandrine},
        title = "{Joint evolution of equatorial oscillation and interhemispheric circulation in Saturn's stratosphere}",
      journal = {Nature Astronomy},
         year = {2022},
        month = {may},
       volume = {6},
        pages = {804-811},
          doi = {10.1038/s41550-022-01670-7}
}

@article{boissinot2024jupiter,
       author = {{Boissinot}, Alexandre and {Spiga}, Aymeric and {Guerlet}, Sandrine and {Cabanes}, Simon and {Bardet}, Deborah},
        title = "{Global climate modeling of the Jupiter troposphere and effect of dry and moist convection on jets}",
      journal = {Astronomy \& Astrophysics},
         year = {2024},
        month = {jul},
       volume = {687},
          eid = {A274},
        pages = {A274},
          doi = {10.1051/0004-6361/202245220}
}

@article{clement2024storms,
  title={Storms and convection on Uranus and Neptune: impact of methane abundance revealed by a 3D cloud-resolving model},
  author={Cl{\'e}ment, No{\'e} and Leconte, J{\'e}r{\'e}my and Spiga, Aymeric and Guerlet, Sandrine and Selsis, Franck and Milcareck, Gwena{\"e}l and Teinturier, Lucas and Cavali{\'e}, Thibault and Moreno, Rapha{\"e}l and Lellouch, Emmanuel and others},
  journal={Astronomy \& Astrophysics},
  volume={690},
  pages={A227},
  year={2024},
  publisher={EDP Sciences},
  doi={10.1051/0004-6361/202348936}
}

@inproceedings{Clement2024ICM,
  author = {Cl{\'e}ment, J.-B. and Forget, F. and Lange, L. and Vos, E. and Millour, E. and Naar, J. and Vandemeulebrouck, R.},
  title = {Investigating Long-Term Mars Climate Evolution: The Planetary Evolution Model},
  booktitle = {Tenth International Conference on Mars},
  year = {2024},
  month = jul,
  address = {Pasadena, United States},
  note = {LPI Contribution No. 3007, id. 3064, HAL: insu-04703198},
  url = {https://insu.hal.science/insu-04703198}
}
\bibliographystyle{aasjournal}

\end{document}